\documentclass[fleqn,usenatbib]{mnras}

\usepackage{newtxtext,newtxmath}
\usepackage[T1]{fontenc}

\DeclareRobustCommand{\VAN}[3]{#2}
\let\VANthebibliography\thebibliography
\def\thebibliography{\DeclareRobustCommand{\VAN}[3]{##3}\VANthebibliography}

\usepackage{graphicx}	% Including figure files
\usepackage{amsmath}	% Advanced maths commands

\usepackage{scalerel}
\usepackage{tikz}
\usetikzlibrary{svg.path}
\usepackage{adjustbox}

\definecolor{orcidlogocol}{HTML}{A6CE39}
\tikzset{
  orcidlogo/.pic={
    \fill[orcidlogocol] svg{M256,128c0,70.7-57.3,128-128,128C57.3,256,0,198.7,0,128C0,57.3,57.3,0,128,0C198.7,0,256,57.3,256,128z};
    \fill[white] svg{M86.3,186.2H70.9V79.1h15.4v48.4V186.2z}
                 svg{M108.9,79.1h41.6c39.6,0,57,28.3,57,53.6c0,27.5-21.5,53.6-56.8,53.6h-41.8V79.1z M124.3,172.4h24.5c34.9,0,42.9-26.5,42.9-39.7c0-21.5-13.7-39.7-43.7-39.7h-23.7V172.4z}
                 svg{M88.7,56.8c0,5.5-4.5,10.1-10.1,10.1c-5.6,0-10.1-4.6-10.1-10.1c0-5.6,4.5-10.1,10.1-10.1C84.2,46.7,88.7,51.3,88.7,56.8z};
  }
}

\newcommand\orcidicon[1]{\href{https://orcid.org/#1}{\mbox{\scalerel*{
\begin{tikzpicture}[yscale=-1,transform shape]
\pic{orcidlogo};
\end{tikzpicture}
}{|}}}}

\title[Image Stacking E+A Galaxies]{Extended structure in E+A galaxies via image stacking}

\author[K. Broadbelt et al.]{
Kieran Broadbelt$^{1}$ \orcidicon{0009-0001-6333-3270},
Kevin Pimbblet$^{1,2}$ \orcidicon{0000-0002-3963-3919},
Daniel J. Farrow$^{1,2}$ \orcidicon{0000-0003-2575-0652},
and Sugata Kaviraj$^{3}$ \orcidicon{0000-0002-5601-575X}
\\
$^{1}$E.A. Milne Centre, Faculty of Science and Engineering, University of Hull, Cottingham Road, Hull HU6 7RX, UK\\
$^{2}$Centre of Excellence for Data Science, Artificial Intelligence and Modelling (DAIM), University of Hull, Cottingham Road, Hull HU6 7RX, UK\\
$^{3}$Centre for Astrophysics Research, Department of Physics, Engineering \& Computer Science, University of Hertfordshire, Hatfield AL10 9AB, UK
}

\date{Accepted XXX. Received YYY; in original form ZZZ}

\pubyear{\the\year{}}

\begin{document}
\label{firstpage}
\pagerange{\pageref{firstpage}--\pageref{lastpage}}
\maketitle

\begin{abstract}
Small-scale interactions such as minor mergers, flybys and harassment are considered common driving mechanisms that can both initiate and quench star-forming processes within post-starburst (PSB) galaxies. PSBs have been linked to disturbed morphologies and faint tidal features in numerous works. We aim to independently study these features in E+A galaxies found in the Galaxy and Mass Assembly Fourth Data Release (GAMA DR4). Using image stacking techniques, we derive the averaged radial surface brightness profiles of E+A image stacks alongside the S\'ersic parameters that best fit these profiles. These E+A stacks consist of 57 images extracted from the Sloan Digital Sky Survey (SDSS) in $g, r$ \& $i$ bands. The E+A stacks are shown to have brighter extended regions than similar non-PSB galaxies, which is indicative of faint tidal features that could be caused by small-scale interactions. Additionally, we make use of Dark Energy Spectroscopic Instrument (DESI) Legacy Survey imaging, which is deeper than SDSS, to further support our conclusions and find that the Legacy Survey images independently strengthen the evidence for excess light at extended radii relative to the comparison samples. We discuss these findings in the context of future surveys such as the Vera C. Rubin Legacy Survey of Space and Time (LSST) survey.
\end{abstract}

\begin{keywords}
galaxies: general -- galaxies: interactions -- galaxies: starburst -- techniques: image processing
\end{keywords}

%%%%%%%%%%%%%%%%%%%%%%%%%%%%%%%%%%%%%%%%%%%%%%%%%%

%%%%%%%%%%%%%%%%% BODY OF PAPER %%%%%%%%%%%%%%%%%%

\section{Introduction}

Post-starburst (PSB) galaxies are considered an important evolutionary link between star-forming spiral galaxies and quiescent elliptical galaxies \citep{DressGunn1983,CouchSharp1987,Wilkinsonetal2017,Pawliketal2018,Greene2021}. They are formed when a rapid starburst is followed by an abrupt halt in star-formation \citep{Zabludoffetal1996}. These rapid quenching systems are thought to account for half of the growth of the red sequence \citep{Belli2019}, they are a well-studied and observed group of galaxies. There have been many physical mechanisms proposed to explain what causes both the burst and quenching of this star formation. Feedback from an active galactic nuclei (AGN) can strip cold gas from a galaxy, halting star-formation \citep{DiMatteo2005, Hopkins2006}. Large galaxy-scale outflows, driven by both supernovae and AGN activity, have been observed to play a role in the quenching process \citep{LiBryan2020, Spilker2020} and have been further studied in hydrodynamical simulations \citep{Nelson2019}. Major mergers have been proposed to both start and halt ongoing star-formation \citep{Springel2005, Wellons2015, Renaud2022}. Additionally, small-scale interaction features are commonly found in PSB galaxies, suggesting flybys or harassment to be pathways for starbursts and quenching \citep{Verrico2023, DOnofrioetal2025}. For the purpose of this paper, we define small-scale interactions as minor mergers, flybys or harassment that result in faint structural features such as tidal arms.

PSB galaxies provide a valuable environment for the study of these numerous mechanisms, with various pathways, including minor mergers, flybys, and internal instabilities, having been proposed to trigger the rapid quenching in PSBs \citep{Pawlik2019}. PSBs are identified by spectra dominated by A-class stars, which is indicative of a major burst of star-formation followed by a sudden and intense quenching event $\sim1$ Gyr ago \citep{Zabludoffetal1996}. \citet{Setton2020} infers that the quenching process likely occurs uniformly throughout the PSB galaxies, and \citet{Verrico2023}, shows that visual classifications indicate that merger features are associated with younger PSBs. Observations suggest that the removal of the cold gas reservoirs comes after the actual quenching process, but the gas reservoirs disappear rapidly within $100-200$ ${\rm{Myr}}$ post-quenching \citep{Suess2017, Bezanson2022}. \citet{Spilker2022} and \citet{DOnofrioetal2025} both present separate discoveries of large molecular gas features in the tidal tail regions of massive $z=0.646$ and $z=0.706$ PSB galaxies. They both suggest the tidal removal of cold molecular gas to be the cause of quenching in the respective galaxies. \citet{DOnofrioetal2025} further discusses the idea that a `radio-mode' (or `maintenance-mode') AGN helps to maintain the quenched state in the context of PSB galaxies. The results conclude that the AGN may have turned on after the starburst phase and is preventing the accretion and cooling of gas. This is supported by previous findings that indicate AGN activity succeeding the initial starburst-driven quenching \citep{Yesuf2014, Pawliketal2018}. 

A specific type of PSB, in particular, we wish to study in this work is an E+A (or k+A) galaxy. They are thought to be a transitional phase between spiral and elliptical galaxies and were first identified by \citet{DressGunn1982}, having elliptical morphologies and stellar populations that are dominated by A-class stars. They display spectra with no $\rm{[OII]}$ forbidden lines but show deep Balmer absorption \citep{Zabludoffetal1996}. More recent works include an additional measure of low $\rm{H\alpha}$ when classifying as E+A \citep{Gotoetal2003, Goto2007b, Wilkinsonetal2017, Pawliketal2018, Greene2021}. A strong $\rm{[OII]}$ equivalent width (EW) measure is an indicator of ongoing star-formation, and the presence of $\rm{H\delta}$ is a sign of recently formed A-class stars. $\rm{H\alpha}$ is often a sign of ongoing star-formation, but it can also be an indicator of AGN activity. When we find the presence of $\rm{H\alpha}$ along with a lack of $\rm{[OII]}$, it can suggest dust obscuration \citep{Gotoetal2003, Pawliketal2018} and as such, many works include the limit on $\rm{H\alpha}$ to reduce this possibility.

In \citet{Broadbelt2025}, we investigated the extremes of the E+A population in GAMA by using an Isolation Forest (iForest). This work resulted in the discovery of E+A galaxies with strong $\rm{H\alpha}$ emission and Balmer absorption with a lack of $\rm{[OII]}$. Both active galactic nuclei (AGN) and dust obscuration were ruled out as a cause for these spectral lines, and as such, we concluded these E+A galaxies were in fact still star-forming. Visual analysis of these galaxies showed that many E+As have close neighbours or faint tidal features that could result in this star formation. Through the use of morphological parameters such as concentration, asymmetry, smoothness, Gini, and ${\rm{M_{20}}}$, we found that few have quantitatively disturbed morphologies. This lack of evidence could be due to the depth of SDSS imaging hiding faint extended structures. As such, in this work, we aim to uncover the extended structure of the E+A galaxies using image stacking techniques to see if small-scale interactions are prevalent in E+A galaxies. 

We utilise image stacking techniques to stack selected E+A galaxies from the Galaxy and Mass Assembly Data Release 4 \citep[GAMA DR4][]{Driveretal2022}. There have been many studies of E+A galaxies, analysing multiple selection methods  \citep{Gotoetal2003, Wilkinsonetal2017, Pawliketal2018, Greene2021}, studying the effects of AGNs and quenching \citep{Kaviraj2007,Yangetal2008,Whitaker2012,Pengpei2025} and studies of E+A environments \citep{Hoggetal2006,Poggiantietal2009, Verganietal2010, Sandoval2025}. We note, however, that there is a lack of studies that aim to probe the extended structure of these E+A galaxies through methods such as image stacking. These methods have been used extensively to measure low surface brightness structures of high-$z$ and faint dwarf galaxies \citep{Lietal2016, Lietal2025, Tianetal2025} as well as being applied to optical \citep{Zibetti2004, vanderwel2008, vandokkum2010}, infrared \citep{Bourne2012,Guo2013} and radio \citep{Garn2009, Hancock2011} images. \citet{vandokkum2010} shows that the average circularised effective radius and the \citet{Sersic} $n$ can both be recovered efficiently from stacked images, with a key advantage being the detection of faint outer regions. Recent works have applied S\'ersic fits to PSB galaxies and found they typically show intermediate morphologies \citep[e.g.][]{Sazonova2021}.

In this work, we will present our findings from the image stacks of 57 E+A galaxies found in GAMA DR4. We will generate comparison stacks using SDSS and Legacy Survey imaging and perform S\'ersic fitting to study the extended structure. In Section \ref{sec:data}, we will discuss in further detail the data we use in selecting the E+A galaxies, the methods used in selecting these galaxies and the code we utilise for the S\'ersic fitting. Section \ref{sec:stacking} describes the image stacking process, including the cutout, masking and convolving of the E+A galaxy images. Section \ref{sec:errors} discusses the generation of the comparison stacks, alongside the error estimation for the core E+A stacks.

%%%%%%%%%%%%%%%%%%%%%%%%%%%%%%%%%%%%%%%%%%%%%%%%%%

\section{Data}
\label{sec:data}

\subsection{GAMA Data and SDSS Images}

We take our data from GAMA DR4 \citep{Driveretal2022} and the necessary Data Management Units (DMUs) required for analysis \citep{Tayloretal2011,Kelvinetal2012,Broughetal2013,Kelvinetal2014, Liskeetal2015,Gordonetal2017,Bellstedtetal2020}. The GAMA DR4 compiles over 250,000 redshifts and, in combination with earlier surveys, results in over 300,000 spectra across five sky regions. GAMA also compiles several data management units (DMUs) that contain a wide variety of additional galaxy properties that are vital for this paper's analysis. The imaging data we use is extracted from SDSS ($ugriz$) using the {\sc{astroquery}} python tool \citep{Astroquery}, and the PSF and magnitude values are taken from the SersicPhotometry DMU \citep{Kelvinetal2012}, calculated primarily using GALFIT and SExtractor. We focus primarily on the bands $gri$. All native reduced frames are downloaded from the SDSS DR7 archives, which are scaled to a uniform zero-point. As part of the SDSS pipeline, the data has already been bias-subtracted and flat-fielded \citep{stoughtonetal2002}. For more information discussing the {\sc{sigma}} pipeline used for calculating the PSF and other imaging data, see \citet{Kelvinetal2012}.

\subsection{Legacy Survey Images}
Additional image stacking is performed with images from the Dark Energy Spectroscopic Instrument (DESI) Legacy Survey Data Release 10 \citep[hereafter the Legacy Surveys][]{DESI}. These images are deeper than the SDSS imaging, which should make the faint substructures more visible and resolved. We search the coordinates of our E+A galaxies and extract the DESI images where possible using the Legacy Surveys cutout service\footnote{\hyperlink{https://www.legacysurvey.org/dr10/description/}{https://www.legacysurvey.org/dr10/description/}}. This extraction results in us losing 3 E+A galaxies where the $i$ band image is missing.

The image stacking process performed on the Legacy Surveys is similar to the E+A stacking process, which will be described in Section. \ref{sec:stacking}, with the main difference being that we use the Legacy Surveys PSF values from the extracted FITS files. Sky background subtraction has been applied to the Legacy Survey DR10. The NOIRLab Community Pipeline removes a sky level by incorporating a sky pattern, an illumination correction, and a fringe pattern correction with the intent to bring the sky level close to zero\footnote{\hyperlink{https://nsf-noirlab.gitlab.io/csdc/csdc-mso-docs/PL206/}{https://nsf-noirlab.gitlab.io/csdc/csdc-mso-docs/PL206/}}. The Legacypipe processing provides additional fitting to remove further sky background \citep{DESI}. Despite these corrections, residual sky patterns may remain in the images. In particular, changes to the DECam Community Pipeline introduced between DR8 and DR9 produced structured sky residuals across all optical bands. A template based sky pattern correction was applied from DR9 onwards and extended into DR10, however, low-level residuals may persist. We discuss the effect and how we solve for this sky background in Section. \ref{sec:skybackground}.

\subsection{Sample Selection}
\label{sec:easelect}
There are a variety of methods used to select E+A galaxies, many of which selecting on low / absent $\rm{[OII]\lambda}3727$ (hereafter as $\rm{[OII]}$) and strong $\rm{H\delta}$ EW \citep{Zabludoffetal1996, Dressler1999}, with other works including additional cuts on $\rm{H\alpha}$ EW \citep[e.g.][]{Goto2005, Goto2007b}, or utilising $\rm{H\alpha}$ and A/K ratios \citep[e.g.][]{Quintero_2004}. There are also selection techniques that use shocked emission line ratios \citep{Alatalo_2016} or AGN-like line ratios \citep{Baron2022}. We utilise the selection techniques discussed in \citet{Wilkinsonetal2017} and commonly used in \citet{Goto2005, Goto2007b}. This selection technique is a cut on $\rm{EW_{[OII]}} < 2.5$\r{A}, $\rm{EW_{H\alpha}} < 3$\r{A} emissions and a cut on $\rm{EW_{H\delta}} < -5$\r{A} absorption. Our selection results in a sample of E+A galaxies with redshift ranging from $0.12 \le z \le 0.4$ and stellar mass ${\rm{}10^8M_\odot \le M_{stellar} \le 10^{11} M_\odot}$

With our E+A galaxies selected, we include an additional selection criteria of signal-to-noise (S/N) $\ge3$. The S/N cut is performed the same as in \citet{Broadbelt2025} and is as follows:
\begin{equation}
    {\rm{S/N}}_{H\delta} = \frac{\rm{|EW|}}{\rm{{EW}_{err}}} \ge 3. 
\end{equation}
This reduces the sample size of `pure' E+A galaxies from 62 to 22, as such, we continue with two samples. Hereafter, the larger stack will be referred to as low S/N E+As, and the reduced stack will be referred to as high S/N E+As.

\subsection{PySersic}
\label{sec:pysersic}
{\sc{pysersic}} is a code for fitting S\'ersic profiles to galaxy images using Bayesian Inference . It is built using the {\sc{jax}} framework with inference performed using the {\sc{numpyro}} probabilistic programming library \citep[][and references therein]{pysersic}. This code was developed to implement S\'ersic fitting in a fully Bayesian context at speed.

%%%%%%%%%%%%%%%%%%%%%%%%%%%%%%%%%%%%%%%%%%%%%%%%%%

\section{Stacking Procedure}
\label{sec:stacking}

\subsection{Residual Sky Background}
\label{sec:skybackground}
Utilising {\sc{astropy}}, we measure the sky background in blank regions of our cutouts for both the SDSS and Legacy Survey images to check for gradients. To do this, we find an initial background estimate and take footprints that are 2$\sigma$ below this level to ensure no sources are being selected. None of the Legacy Survey images extracted have gradient or sky level flags and have average residual sky levels of $0.0003\pm0.0002$ nMgy pixel$^{-1}$, as such, no further sky corrections are applied to these images. For the SDSS images, we find an average sky level of $0.003\pm0.0026$ nMgy pixel$^{-1}$. 6 SDSS images are flagged with sky levels $\sim1$dex above this average but these are all removed due to either contamination and manual removal or the PSF constraints. We therefore perform no further sky correction on the SDSS images.

\subsection{Image Stacking}
To investigate the surface brightness profiles of the two E+A samples, we perform stacking analysis on single-band images for each sample separately. We will discuss the process of this stacking below.
\label{sec:imagestack}

    \subsubsection{Image Cut-out Extraction}
    \label{sec:extraction}
    Given the RA and DEC (deg) coordinates of our galaxy sample, we extract cut-out images centred on the coordinates of the desired object. We utilise the Python tool {\sc{astroquery}} for extracting the images from the SDSS archive. These cut-out images have a postage stamp size of $50\times50$ pixels, corresponding to $20''\times20''$. We then mask our cut-out images by applying an elliptical patch to the image which is twice the minor and major axis extracted from the ApMatchedPhotom DMU \citep{Driveretal2016}. The elliptical mask creates a `clean' zone that is not affected by the later thresholding, we do this to preserve any faint extended structure. To remove residuals and large contaminations caught in the cut-out we apply a detection threshold. The threshold removes objects $1.5-3\sigma$ above the background, with a clean zone of the aforementioned  $2\times r_{\rm{minor}}, r_{\rm{major}}$ elliptical patch. Unfortunately, several objects are found that still contain residual flux, notably multiple nuclei systems or bright foreground objects overlapping the E+A galaxy, we remove these objects from the sample, reducing the low S/N sample to 58 galaxies and the high S/N sample to 18 galaxies. 
    
    We find that the average axis ratio for our E+A galaxies is $r_{\rm{minor}}/r_{\rm{major}}\approx0.8$, with the lowest ratio being $0.4$. Since the stacked position angles are randomised and the median systems are not strongly edge-on, circularised radial profiles provide an adequate first-order measure of the average surface-brightness structure.
    
    \subsubsection{Redshift Matching}
    We scale all of the images to a rest frame of $z = 0.12$. This redshift is selected as it is the lowest redshift in the sample data and a non-zero value prevents zero-division errors. We make use of the Python package {\sc{astropy}} \citep{astropy:2022} and its tool Planck18 \citep{Planck2018} to scale all of the cut-outs to the same redshift and then crop the enlarged images back down to the $50\times50$ pixel square for stacking. Prior to the final stacking, we normalise each galaxy image based on total flux, which resolves the problem of cosmological redshift surface brightness dimming that would occur due to the rescaling.
    
    \subsubsection{PSF Matching and Convolution}
    Additional matches on the point spread function (PSF) are performed to ensure all of the images are convolved to the same PSF. Before we convolve, we set a limit on the PSF and remove any images with a PSF value $\ge 1.45''$, which reduces the sample size to 57 in the low S/N E+A stack and 17 images in the high S/N E+A stack. The PSF limit is to ensure the results aren't muddied by extreme PSF wings. We make use of the Python package {\sc{photutils}} \citep{photutils} and its tool make\_2dgaussian\_kernel and create\_matching\_kernel. This tool ensures that the convolutions match the desired PSF. The PSF value we convolve to is the maximum PSF of the sample which can be no higher than $1.45''$. The S\'ersic profile we create later is also convolved to this PSF value. Because of the greater resolution, the maximum PSF for the Legacy Survey images is set to $1.00''$

    \subsubsection{Final Stacking}
    The image stacking process is performed by {\sc{imcombine}} \footnote{\url{https://astropop.readthedocs.io/en/latest/reduction/imarith.html}}, which is a tool in the {\sc{astropop}} Python library \citep{astropop}. We use its functionality to ignore the NaN values that result from the object masking. We also stack our images using `mean' stacking to preserve the total flux and faint outer light better. We create stacks from the two E+A samples, consisting of 57 images and 17 images respectively. 
    
    \subsubsection{Surface Brightness Profile and S\'ersic Fitting}
    \label{sec:sersic}
    Radial profiles of the stacks are generated using circular apertures, covering $0.01\to 20$ ${\rm{pix}}$ in steps of 1pix. The total flux inside each aperture range is calculated and a surface brightness profile is generated. This flux measurement is in nanomaggies, typical of SDSS images.
    
    {\sc{pysersic}} is used to generate the S\'ersic profile fitting parameters, using the definitions from \citet{Sersic}. The empirically derived S\'ersic profile is the most common parametric form for the surface-brightness profile and can be defined as:
    \begin{equation}
        \label{eq:sersic}
        I(r) = I_{\rm{e}} \exp{\left\{-b_n\left[\left(\frac{r}{r_e}\right)^{1/n}-1\right]\right\}},
    \end{equation}
    where $n$ is known as the S\'ersic index, $b_n$ is a coefficient chosen so that a circular aperture with radius $r_e$ contains half of the galaxy's flux, and $I_e$ is the surface brightness at $r=r_e$. There are several special cases for the S\'ersic index, namely $n=1$, corresponding to an exponential profile and $n=4$  corresponding to a de Vaucouleurs profile \citep{Vaucouleur}. Equation (\ref{eq:sersic}) defines a 1D profile. {\sc{pysersic}} also includes the parameters required to represent a 2D profile, these parameters are; the centre of the profile ($x_c, y_c$), the ellipticity and the orientation of the elliptical isophotes. For our work, we will only be considering the 1D profile.

\begin{figure}
    \centering
    \includegraphics[width=\columnwidth]{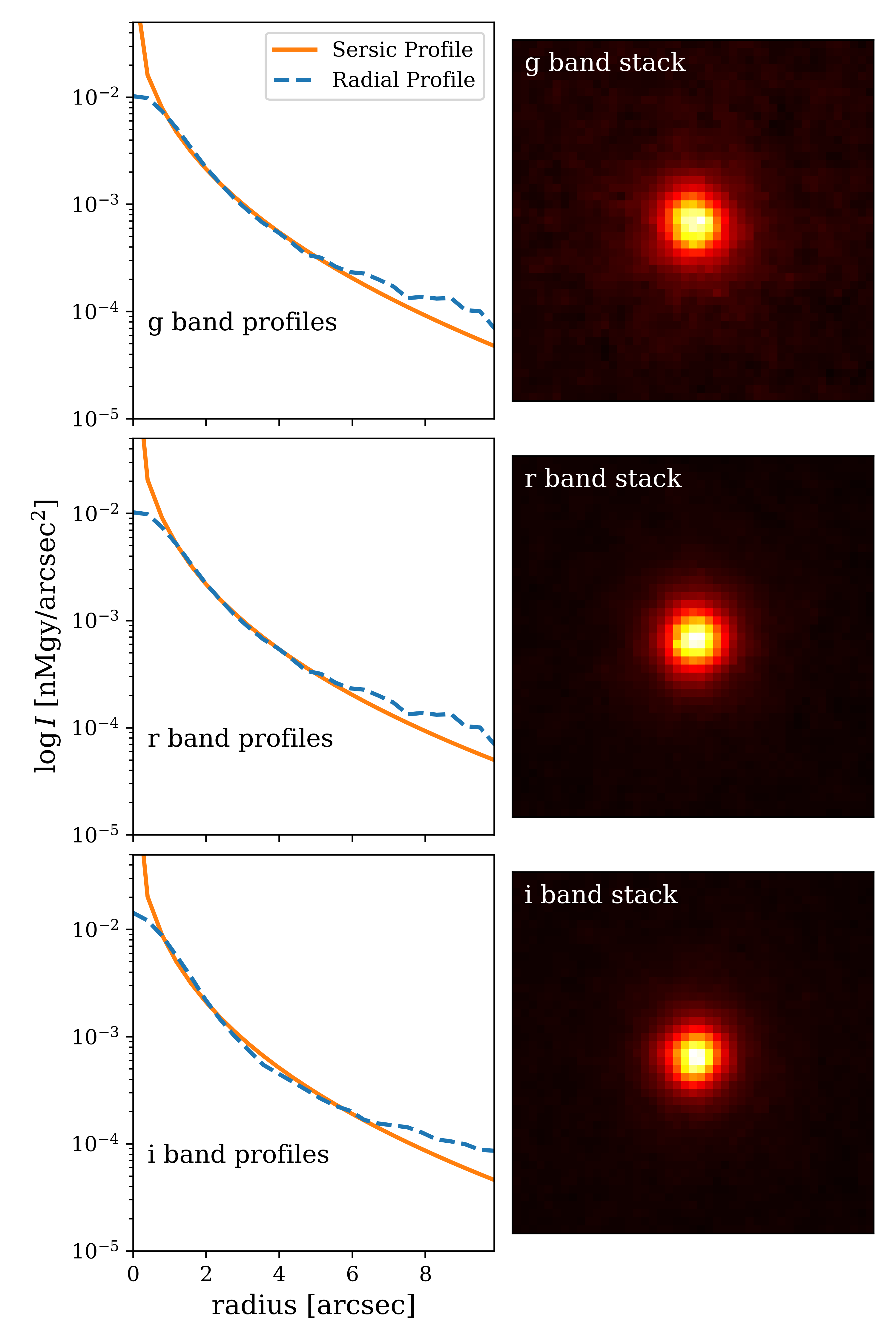}
    \caption{Surface brightness profile of the SDSS low S/N E+A stack (left) and the stacked image plot (right). The dashed blue line is the radial profile of the surface brightness fitted by circular bins about the central pixel. The solid orange line indicates the fitted S\'ersic profile with the $I_e, r_e$ and $n$ values extracted via {\sc{pysersic}}. Top; $g$ band, middle; $r$ band , and bottom; $i$ band.}
    \label{fig:70stackradprofile}
\end{figure}

\begin{figure}
    \centering
    \includegraphics[width=\columnwidth]{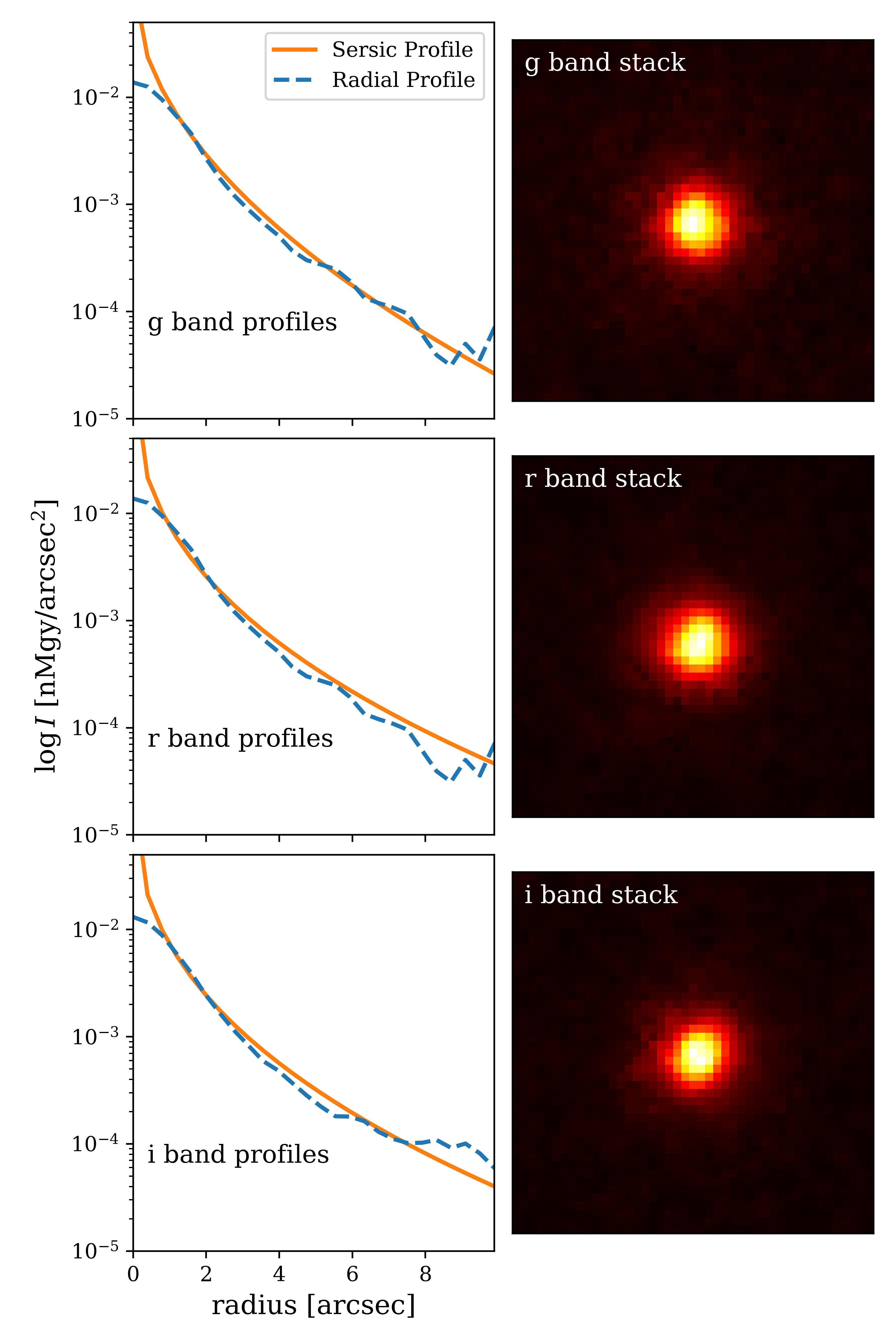}
    \caption{Surface brightness profile of the SDSS high S/N E+A stack (left) and the stacked image plot (right). Line colours and styles are the same as stated in Fig. \ref{fig:70stackradprofile} Top; $g$ band, middle; $r$ band , and bottom; $i$ band.}
    \label{fig:20stackradprofile}
\end{figure}

    \citet{pysersic} utilise the {\sc{jax}} library to drastically decrease computational runtimes to fit S\'ersic profiles. They use a maximum a-posteriori (MAP) approach where the point estimate corresponds to the MAP or `best-fit' parameter values. This means it can produce full posterior distributions using multiple methods. We make use of the Markov chain Monte Carlo (MCMC) approach to sampling the S\'ersic parameters, as this is the most robust and accurate method. The fitting model requires five major inputs, i) the cutout image of the galaxy, for us this is the processed stack, ii) a weightmap that is the same shape as the image, for this we generate 200 randomly rotated stacks and calculate the standard deviation at each pixel, iii) a prior which is initialized by {\sc{pysersic}} before the fitting which for our case is a `S\'ersic' prior with `none' sky type, iv) an optional mask which we don't use as the {\sc{imcombine}} stacking function already masks `bad' pixels, and v) a loss function, we use the cash\_loss function. The Cash loss utilises a Cash statistic based on Poisson statistics derived in \citet{CashLoss} and advocated for in \citet{Erwin2015} for use in S\'ersic fitting. We use the `none' sky type here as the negative values associated with different sky models, such as `flat' or `tilted-plane', will cause issues with the Cash loss function. Figs. \ref{fig:70stackradprofile} \& \ref{fig:20stackradprofile} display the S\'ersic profiles generated from the {\sc{pysersic}} results and the surface brightness profiles generated from the radial bins. They show both profiles alongside the image stacks in the $gri$ bands. In both the low and high S/N stacks, the S\'ersic profiles begin to deviate and underpredict at larger radii.

%%%%%%%%%%%%%%%%%%%%%%%%%%%%%%%%%%%%%%%%%%%%%%%%%%

\subsection{Comparison Stacks}
\label{sec:errors}
The extended structure of galaxies can be strongly affected by contamination. We have discussed the masking process in Section \ref{sec:extraction} and any galaxies that were too contaminated we dropped from the dataset. We apply two tests to evaluate the robustness of our masking and stacking procedure: (i) a jackknife analysis of the entire stacking process; (ii) a rotation analysis where a random angle of rotation is applied to each image before the masking and stacking procedure. 

We further create comparison stacks to compare to the E+A stacks, generating radial and S\'ersic profiles in the same manner discussed in Section \ref{sec:imagestack}.

    \subsubsection{E+A Analysis Stacks}
    \label{sec:rotatestack}
    We use the jackknife analysis to estimate the uncertainty introduced by individual galaxies that may disproportionately influence the stacked profiles. For each iteration, a single galaxy image is excluded from the sample prior to stacking, after which the full {\sc{pysersic}} fitting pipeline is applied. The resulting ensemble of stacked profiles is then used to estimate the variance associated with sample selection and potential outliers.
    
    The extended structure of galaxies can be significantly affected by contamination and residual alignment effects. To ensure the stacked profiles provide a robust measure of the average surface-brightness profile, we apply a series of random rotations and reflections to each image before stacking. Owing to the finite pixel resolution of the images, arbitrary rotations (e.g. $45^\circ$) require interpolation and can redistribute flux between pixels, potentially introducing systematic artefacts. Rotations are therefore restricted to multiples of $90^\circ$, with random horizontal and vertical reflections also applied. These transformations preserve the flux distribution while randomising the orientation of any residual asymmetric features, such that the resulting stacked profile is expected to be rotationally symmetric and the azimuthally averaged radial profile an appropriate representation of the mean galaxy structure \citep{Lietal2016}.

    The uncertainty on the stacked surface-brightness profiles is estimated from the distribution of profiles obtained from the jackknife and random-rotation stacks. At each radial bin, the standard error is calculated from the scatter between realisations, providing a measure of the sensitivity of the stacked profile to sample selection and image orientation.
    
    Given the small size of the high-S/N sample, we additionally perform a bootstrap resampling analysis. For each of 1000 iterations, a new sample is constructed by drawing galaxies from the E+A sample with replacement until the original sample size is recovered. The images are processed and stacked using the same procedure applied to the primary sample, and a radial surface-brightness profile is measured from the resulting stack. The standard errors derived from this distribution of stacked profiles are propagated through subsequent analyses and are shown combined with the jackknife error envelopes on the comparative radial surface-brightness profiles.
    
    \subsubsection{Quiescent and Star-forming Comparisons}
    \label{subsec:matching}

    \begin{figure}
        \includegraphics[width=\columnwidth]{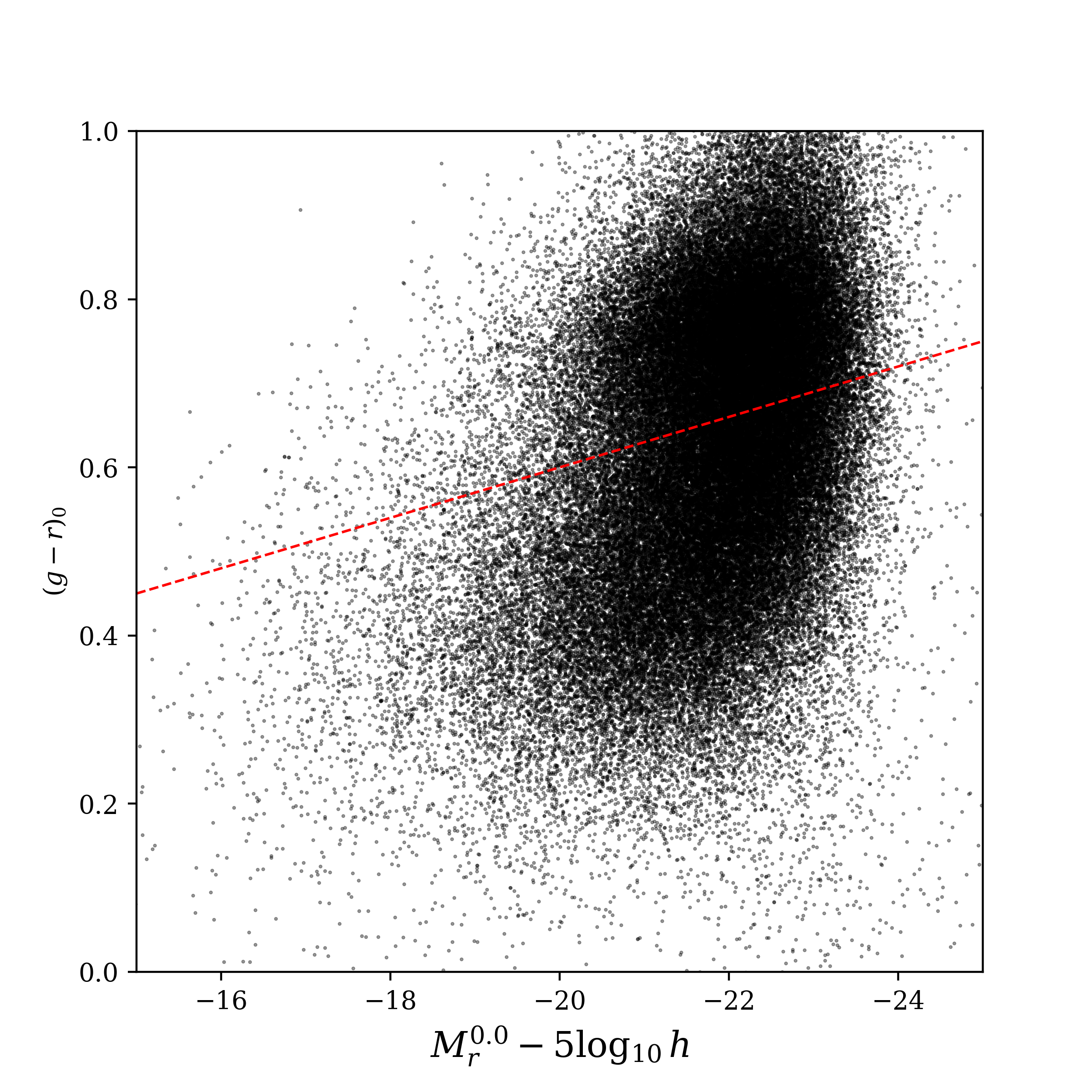}
        \caption{Restframe $g-r$ colour vs Petrosian absolute $r$-band magnitude diagram used to select for blue star-formers and red quiescents. We use the same delimiting line as in \citet{Farrow2015} and is shown here as a red dashed line.  The plot shows the magnitude limited GAMA survey where $r_{{\rm{petro}}} <19.4$.}
        \label{fig:colourmag}
    \end{figure}

    \begin{figure*}
        \includegraphics[width=\textwidth]{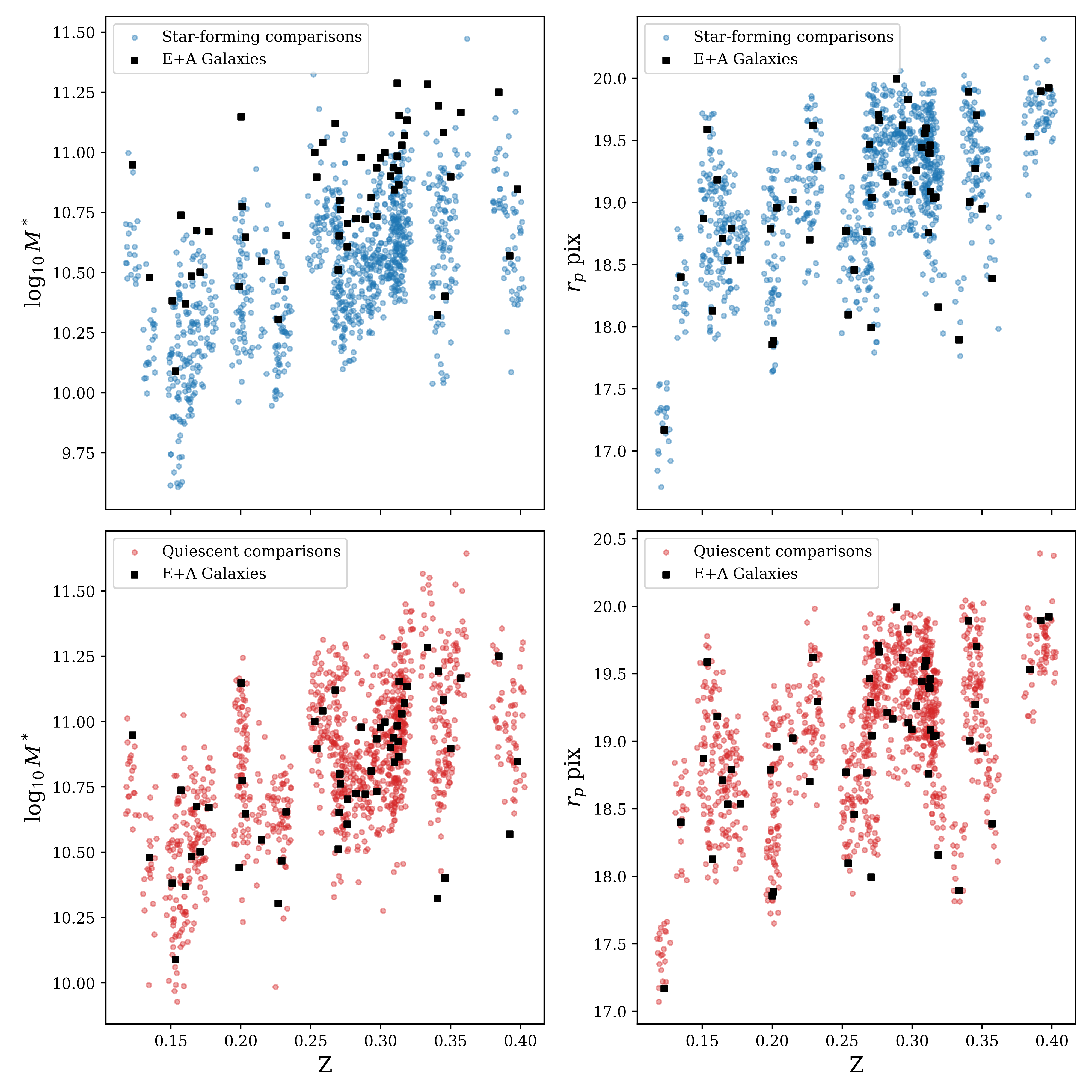}
        \caption{Feature spread of the comparison star-forming and quiescent galaxies.The coloured circles represent the star-forming and quiescent galaxies and the solid black squares, the E+A galaxies. Top plot displays the star-forming galaxies, bottom the quiescent galaxies, left displays the spread of $M^*$ vs $z$ and right $r_p$ vs $z$.}
        \label{fig:featurespread}
    \end{figure*}
    
    To compare the E+A stacks, we create several image stacks from samples of quiescent and star-forming galaxies. We utilise the colour-magnitude diagram from \citet{Farrow2015}, using $g-r$ colour, kcorrected to $z=0$, and the Petrosian absolute $r$-band magnitude, hereafter $M_{r,h}$. To compute $M_{r,h}$ we use the formula,
    \begin{equation}
        M_{r,h} = m_r -k_r(z)+Q(z-z_{{\rm{ref}}})-5{\rm{log_{10}}}(D_L(z))-25,
    \end{equation}
    where $Q$ is the luminosity evolution parameter, $D_L$ the luminosity distance and $z_{ref}$ is the reference redshift, here we adopt $z_{ref}=0$. We adopt a luminosity evolution parameter of $Q=1.45$ \citep{Loveday2015}. We also make use of the kCorrections DMU for our $(g-r)_0$ colour calculations, these kcorrections are fit in each SDSS $ugriz$ band with model magnitudes and SED templates \citep[see in][]{Loveday2012}. Galaxies are divided using a sloping cut with a gradient of 0.03, following typical literature values and \citep{Farrow2015},
    \begin{equation}
        (g-r)_0=-0.030(M_{r,h}-M^*_{r,h})-0.0678.
    \end{equation}
    The population split is shown in Figure. \ref{fig:colourmag}. We take red and therefore quiescent galaxies to be above the sloped cut, and blue star-forming galaxies to be below the sloped cut.

    From the quiescent and star-forming samples, we create a selection of similar galaxies to our E+A galaxies, matched on $z$, stellar mass $M_*$ and Petrosian radius $r_p$. The values for $M_*$ and $r_p$ are found in GAMA DMUs, StellarMasses \citep{Tayloretal2011} and ApMatchedCat \citep{Liskeetal2015}. We find 20 galaxies per E+A galaxy that falls within the range of $z\pm0.005$, $r_p\pm1{\rm{pix}}$, and $\log_{10} M_*\pm 0.5 {\rm{M}}_\odot$. This results in a sample of 1102 quiescent galaxies and 1031 star-forming galaxies. We find that all objects have at least one matched galaxy that falls within the specified range, but not all have 20 matches, resulting in a reduced sample. Through manual visual checks, we search for any artefacts or bad images to remove from the samples, reducing the quiescent sample to 989 and the star-forming sample to 897. Each E+A galaxy still has at least 4 matched quiescent and star-forming galaxies. The spread of features is shown in Figure. \ref{fig:featurespread}, we can see that the star-forming $M_*$ is typically lower than that of the E+A galaxies and is where we find fewer matched galaxies within the specified range. This may be due to star-forming galaxies typically having lower stellar masses than quiescent galaxies \citep{Kauffmann2003, Baldry2006}. We again check the axis ratios of our comparison galaxies to check for edge on objects and find the average axis ratio is $r_{\rm{minor}}/r_{\rm{major}}\approx0.78$ for the star-formers, with a low of $0.5$, and $r_{\rm{minor}}/r_{\rm{major}}\approx0.76$ for the quiescents, with a low of $0.4$. As such, we again argue that the circularised radial profiles provide adequate measures of the surface-brightness structure.

    Utilising the quiescent and star-forming samples, we make stacks of 57 randomly selected galaxies that match the individual E+A galaxies. Additional stacks are made for the high S/N E+A sample, consisting of 17 galaxies that are pulled from the prior quiescent and star-forming samples. The image stacking pipeline is then applied in the same manner as in Section. \ref{sec:stacking} to create 30 quiescent and 30 star-forming galaxy stacks per E+A sample. A total of 120 stacks are therefore generated: 60 high S/N matched stacks and 60 low S/N matched stacks. The surface brightness profiles are calculated for each stack, and the S\'ersic parameter calculations are also performed to find $I_e,r_e$, and $n$.

    For quantising the outer profile difference between the E+A stacks and the comparison stacks, we define an outer profile statistic,
    \begin{equation}
        f_{\rm{outer}}=\frac{F(r>r_2)}{F(r_1<r\le r_2)},
    \end{equation}
    where $F$ is the weighted surface brightness profile within the specified radial range, $r_1$ and $r_2$ define an intermediate reference annulus, here we use $r_1 =5''$ and $r_2=8''$ with $r_{\rm{max}}=10''$. We then compute the difference as,
    \begin{equation}
        \Delta f_{\rm{outer}} = f_{\rm{outer}}^{\rm{E+A}} - f_{\rm{outer}}^{\rm{comp}},
    \end{equation}
    where $f_{\rm{outer}}^{\rm{comp}}$ is measured from either the quiescent or star-forming comparison stacks. To assess the robustness of the measurement, we generated 10,000 bootstrap realisations of the comparison sample by resampling galaxies with replacement and recomputing $f_{\rm{outer}}$ for each realisation. The observed E+A stack is held fixed for these realisations. We derive error measurements and $\sigma$ values. This provides a direct measure of how consistently the E+A stacks exhibit enhanced outer light relative to the comparison samples.

%%%%%%%%%%%%%%%%%%%%%%%%%%%%%%%%%%%%%%%%%%%%%%%%%%

\section{Jackknife and Rotation Results}

The E+A stacks display intermediate S\'ersic indices, with $n_g =2.49\pm0.40$, $n_r =3.02\pm0.40$, and $n_i =3.01\pm0.53$ for the low S/N stack, shown in Table. \ref{tab:lowSNsersic results}. These values of $n\approx3$ are consistent with discy elliptical galaxies and indicate the low S/N stack is likely bulge-dominated. In the high S/N results shown in Table. \ref{tab:highSNsersic results}, we find that $n$ is much lower, being closer to $n=2$, indicating a more transitional, intermediate morphology. The higher S\'ersic indices in the low S/N stack compared to the high S/N stack is a likely result from stacking more images which, while revealing faint structures at larger radii, could also artificially increase the bulge.

For both the low and high S/N stacks we generate 50 different stacks in which the images are transformed as stated in Section. \ref{sec:rotatestack}. We find the averaged profile is rotationally symmetric and the radial profile derived from the stack is therefore sufficient to characterise the averaged profile \citet{Lietal2016}. The random distribution of the position angle has an insignificant effect on the surface brightness profile. Using {\sc{pysersic}} we quantise the S\'ersic parameters to analytically check for differences in the brightness profiles. The results of this analysis are shown in Tables. \ref{tab:lowSNstack} \& \ref{tab:highSNstack} and we can see again that there is an insignificant difference between the rotated stacks and the standard non-transformed stacks, again confirming that the stack is sufficient to characterise the averaged profile.

We next performed jackknife analysis on both the low and high S/N stacks, generating 57 low S/N image stacks and 17 high S/N image stacks. The results of the {\sc{pysersic}} analysis can be found in Tables. \ref{tab:lowSNstack} \& \ref{tab:highSNstack}. The jackknife analysis is similar to the rotation analysis, with both the surface brightness profiles and the S\'ersic parameters showing an insignificant difference. This further supports that the final image stack is reliable and is not overly dependent on individual images. For the high S/N stack, the difference is greater but we consider it to be insignificant due to the low sample size of this image stack resulting in individual images having more weighting than the same image in the low S/N stack.

%%%%%%%%%%%%%%%%%%%%%%%%%%%%%%%%%%%%%%%%%%%%%%%%%%%%%%%%%%%%%%%%%%%%%%%%%%%%%%%%%%%
%%%%%%%%%%%%%%%%%%%%%%%%%% Radial Profile Figures %%%%%%%%%%%%%%%%%%%%%%%%%%%%%%%%%

\begin{table}
\caption{S\'ersic fitting parameters for the SDSS low S/N stack. $r_e$ is the effective radius, $I_e$ is the surface brightness at $r=r_e$ and $n$ is the S\'ersic index. The table shows that both the jackknife and rotated analysis stacks are consistent with the standard low S/N stack.}
\label{tab:lowSNstack}
\begin{tabular}{llll}
\hline
 & Stack Value & Jackknife Mean & Rotation Mean \\ \hline
$g$ band &  &  &  \\ 
$r_e$ [arcsec] & 2.545 & 2.543 & 2.557 \\
$I_e$ $\rm[nMgy/arcsec^2]$ & 0.126 & 0.126 & 0.126 \\
$n$ & 2.493 & 2.498 & 2.417 \\ \hline
$r$ band &  &  &  \\ 
$r_e$ [arcsec] & 2.354 & 2.354 & 2.319 \\
$I_e$ $\rm[nMgy/arsec^2]$ & 0.440 & 0.441 & 0.463 \\
$n$ & 3.019 & 3.008 & 3.016 \\ \hline
$i$ band &  &  &  \\ 
$r_e$ [arcsec] & 2.277 & 2.277 & 2.283 \\
$I_e$ $\rm[nMgy/arsec^2]$ & 0.797 & 0.73 & 0.791 \\
$n$ & 3.005 & 3.009 & 2.999 \\ \hline
\end{tabular}
\end{table}

\begin{table}
\caption{S\'ersic fitting parameters for the SDSS high S/N stack. $r_e$ is the effective radius, $I_e$ is the surface brightness at $r=r_e$ and $n$ is the S\'ersic index. The table shows that both the jackknife and rotated analysis stacks are consistent with the standard high S/N stack.}
\label{tab:highSNstack}
\begin{tabular}{lllll}
 \hline
 & Stack Value & Jackknife Mean & Rotation Mean \\ \hline
$g$ band &  &  &  \\
$r_e$ [arcsec] & 1.852 & 1.875 & 1.894 \\
$I_e$ $\rm[nMgy/arsec^2]$ & 0.551 & 0.535 & 0.528 \\
$n$ & 1.967 & 1.977 & 1.891 \\ \hline
$r$ band &  &  &  \\
$r_e$ [arcsec] & 2.228 & 2.253 & 2.282 \\
$I_e$ $\rm[nMgy/arsec^2]$ & 1.019 & 0.975 & 0.88 \\
$n$ & 2.463 & 2.537 & 2.61 \\ \hline
$i$ band &  &  &  \\ 
$r_e$ [arcsec] & 2.122 & 2.133 & 2.078 \\
$I_e$ $\rm[nMgy/arsec^2]$ & 1.477 & 1.462 & 1.566 \\
$n$ & 2.452 & 2.489 & 2.429 \\ \hline
\end{tabular}
\end{table}

%%%%%%%%%%%%%%%%%%%%%%%%%%%%%%%%%%%%%%%%%%%%%%%%%%%%%%%%%%%%%%%%%%%%%%%%%%%%%%%%%%%

%%%%%%%%%%%%%%%%%%%%%%%%%%%%%%%%%%%%%%%%%%%%%%%%%%
\section{Comparison Stacks}

\subsection{SDSS Images}

\begin{table}
\caption{S\'ersic fitting parameters for the SDSS low S/N E+A stack compared to the comparison Quiescent (Q) and Star-forming (SF). $r_e$ is the effective radius, $I_e$ is the surface brightness at $r=r_e$ and $n$ is the S\'ersic index. The error columns are found via bootstrapping 1000 times.}
\label{tab:lowSNsersic results}
\begin{tabular}{llll}
 \hline
 & E+A Value & Q Value & SF Value \\ \hline
$g$ band & & & \\
$r_e {\rm{[arcsec]}}$        & 2.545 $\pm$ 0.175 & 2.562$\pm$0.125 & 2.757$\pm$0.121 \\
$I_e {\rm{[nMgy/arcsec^2]}}$ & 0.126 $\pm$ 0.024 & 0.140$\pm$0.019 & 0.191$\pm$0.029 \\
$n$                          & 2.493 $\pm$ 0.403 & 1.897$\pm$0.198 & 1.576$\pm$0.172 \\ \hline
$r$ band & & & \\
$r_e {\rm{[arcsec]}}$        & 2.354 $\pm$ 0.174 & 2.264$\pm$0.098 & 2.499$\pm$0.129 \\
$I_e {\rm{[nMgy/arcsec^2]}}$ & 0.444 $\pm$ 0.113 & 0.552$\pm$0.071 & 0.480$\pm$0.047 \\
$n$                          & 3.019 $\pm$ 0.396 & 2.278$\pm$0.165 & 1.692$\pm$0.112 \\ \hline
$i$ band & & & \\ 
$r_e {\rm{[arcsec]}}$        & 2.277 $\pm$ 0.140 & 2.232$\pm$0.103 & 2.575$\pm$0.164 \\
$I_e {\rm{[nMgy/arcsec^2]}}$ & 0.796 $\pm$ 0.129 & 0.921$\pm$0.116 & 0.636$\pm$0.084 \\
$n$                          & 3.005 $\pm$ 0.528 & 2.313$\pm$0.171 & 1.896$\pm$0.124 \\ \hline
\end{tabular}
\end{table}

\begin{table}
\caption{S\'ersic fitting parameters for the SDSS high S/N E+A stack compared to the comparison Quiescent (Q) and Star-forming (SF). $r_e$ is the effective radius, $I_e$ is the surface brightness at $r=r_e$ and $n$ is the S\'ersic index. The error columns are found via bootstrapping 1000 times.}
\label{tab:highSNsersic results}
\begin{tabular}{llll}
 \hline
 & E+A Value & Q Value & SF Value \\ \hline
$g$ band & & & \\
$r_e {\rm{[arcsec]}}$        & 1.852 $\pm$ 0.105 & 2.238$\pm$0.156 & 2.727$\pm$0.216 \\
$I_e {\rm{[nMgy/arcsec^2]}}$ & 0.551 $\pm$ 0.258 & 0.301$\pm$0.053 & 0.303$\pm$0.060 \\
$n$                          & 1.967 $\pm$ 0.705 & 1.938$\pm$0.277 & 1.337$\pm$0.133 \\ \hline
$r$ band & & & \\
$r_e {\rm{[arcsec]}}$        & 2.228 $\pm$ 0.215 & 2.100$\pm$0.120 & 2.539$\pm$0.164 \\
$I_e {\rm{[nMgy/arcsec^2]}}$ & 1.019 $\pm$ 0.721 & 0.954$\pm$0.144 & 0.672$\pm$0.107 \\
$n$                          & 2.463 $\pm$ 0.478 & 2.028$\pm$0.225 & 1.395$\pm$0.134 \\ \hline
$i$ band & & & \\ 
$r_e {\rm{[arcsec]}}$        & 2.122 $\pm$ 0.149 & 2.121$\pm$0.125 & 2.516$\pm$0.188 \\
$I_e {\rm{[nMgy/arcsec^2]}}$ & 1.477 $\pm$ 0.386 & 1.454$\pm$0.241 & 0.966$\pm$0.167 \\
$n$                          & 2.452 $\pm$ 0.679 & 2.228$\pm$0.269 & 1.525$\pm$0.160 \\ \hline
\end{tabular}
\end{table}

\begin{figure*}
    \centering
    \includegraphics[width=\textwidth]{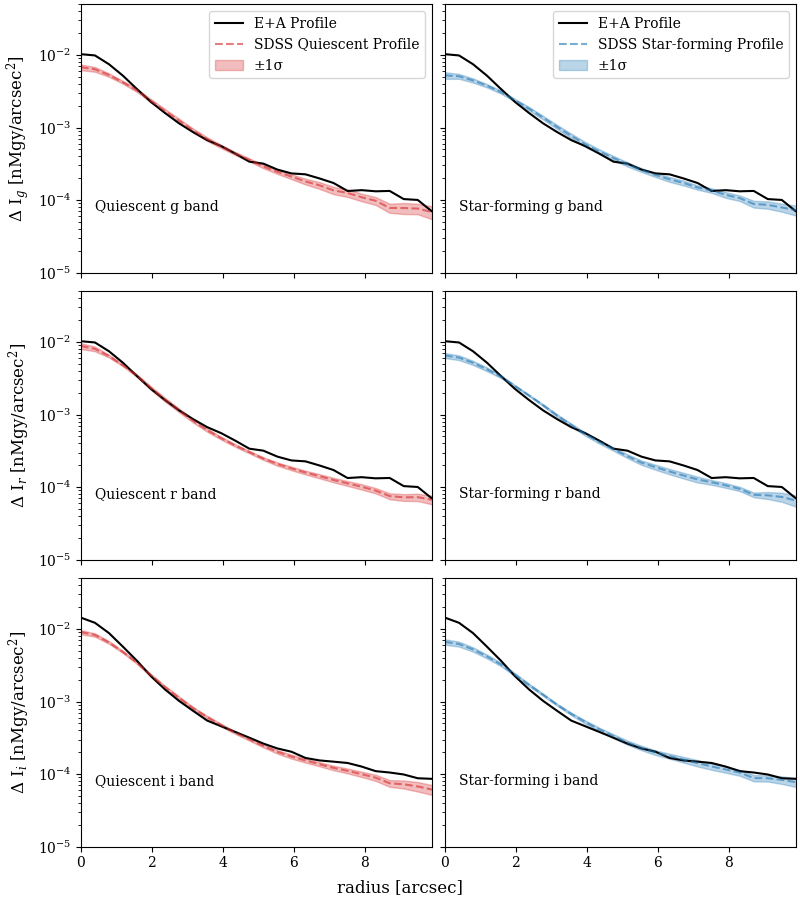}
    \caption{Surface brightness profiles of the low S/N E+A stacks compared to the profiles of the quiescent (left) and star-forming (right) stacks. The solid black line shows the E+A surface brightness profile, with the coloured dashed line representing the comparison mean and the shaded regions representing the comparison spread. Almost all of the E+A stacks show brighter profiles across all radii, with a dip in the middle that roughly matches the quiescent and star-forming stacks. We see the greatest difference at large radii ($r>5''$). The star-forming $g$ band is the only profile that is consistently brighter than the E+A stack, including at greater radii.}
    \label{fig:70comparestacks}
\end{figure*}

\begin{figure*}
    \centering
    \includegraphics[width=\textwidth]{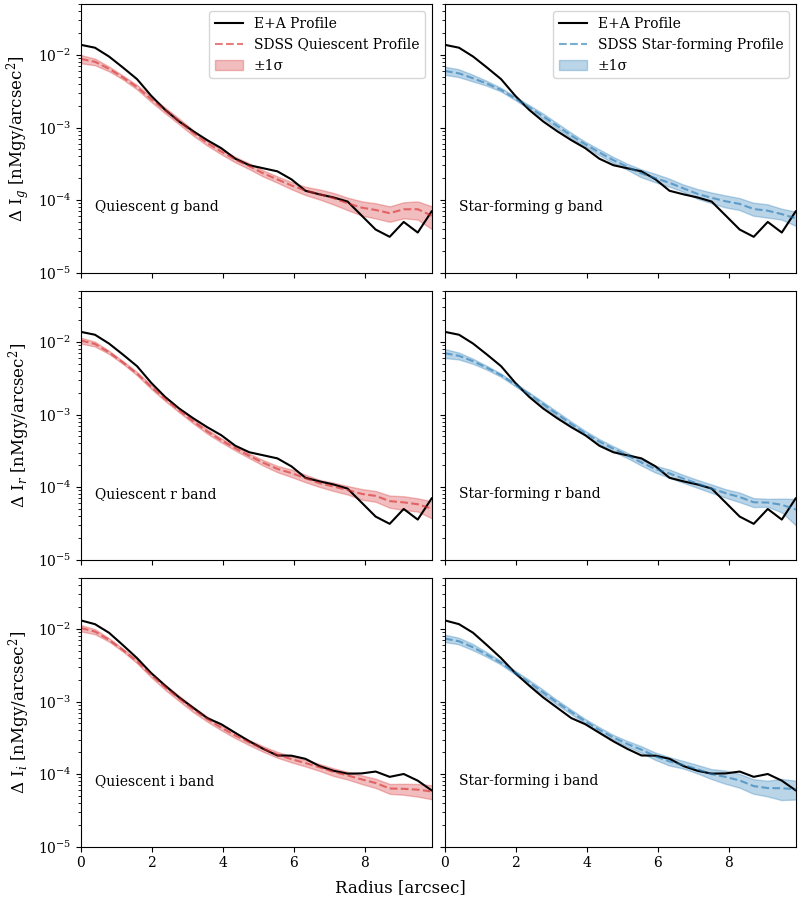}
    \caption{Surface brightness profiles of the high S/N E+A stacks compared to the profiles of the quiescent (left) and star-forming (right) stacks. The key is the same as in Figure. \ref{fig:70comparestacks} The $r$ and $i$ band E+A stacks show brighter profiles across all radii, with the greatest difference seen at large radii ($r>5''$). The $g$ band profiles for both quiescent and star-forming stacks are brighter than the E+A stack at large radii, with the star-forming stack being consistently shallower.}
    \label{fig:20comparestacks}
\end{figure*}

The surface brightness profiles in the low S/N stacks are shallower at larger radii compared to the quiescent and star-forming stacks (Figure. \ref{fig:70comparestacks}), indicating that the E+A population retains a more extended distribution of faint stellar light. In the central regions ($r<2''$) the stronger intensity of the E+A stacks suggest a more concentrated stellar component, consistent with a centrally enhanced recent star formation or a compact PSB remnant \citep{Yang2004, Yangetal2008, French2021}. At intermediate radii ($2''<r<5''$), the higher brightness in the quiescent and star-forming stacks may indicate a smoothly distributed stellar envelopes in these populations. However, at larger radii ($r>5''$), the E+A stacks again become brighter, suggesting the presence of extended, low-surface-brightness components that is less prominent in the comparison samples. This is suggestive of possible merger or interaction features during the PSB phase, consistent with tidal debris or the fading of recently quenched star-forming regions \citep{Pawlik2019}.

Within the high S/N stack we find broadly consistent behaviour, all three $gri$ bands have a more concentrated central region, suggesting a more centrally concentrated evolution event \citep{Yan2019}. The E+A profiles at intermediate radii are again systematically fainter than both the quiescent and star-forming stacks, reinforcing the interpretation that the bulk of the recent stellar mass assembly is confined to the inner regions, with reduced intermediate radius contribution. However, the behaviour at large radii diverges more strongly between bands. Only the $i$ band E+A stack shows enhanced surface brightness relative to the comparisons, while in the $g$ band both quiescent and star-forming galaxies remain brighter, and in the $r$ band, the profiles are broadly consistent. This divergence suggests that the extended structure found in these E+A stacks is either weak or dominated by low surface brightness components that are not robustly traced in all bands.

\begin{figure}
    \centering
    \includegraphics[width=\columnwidth]{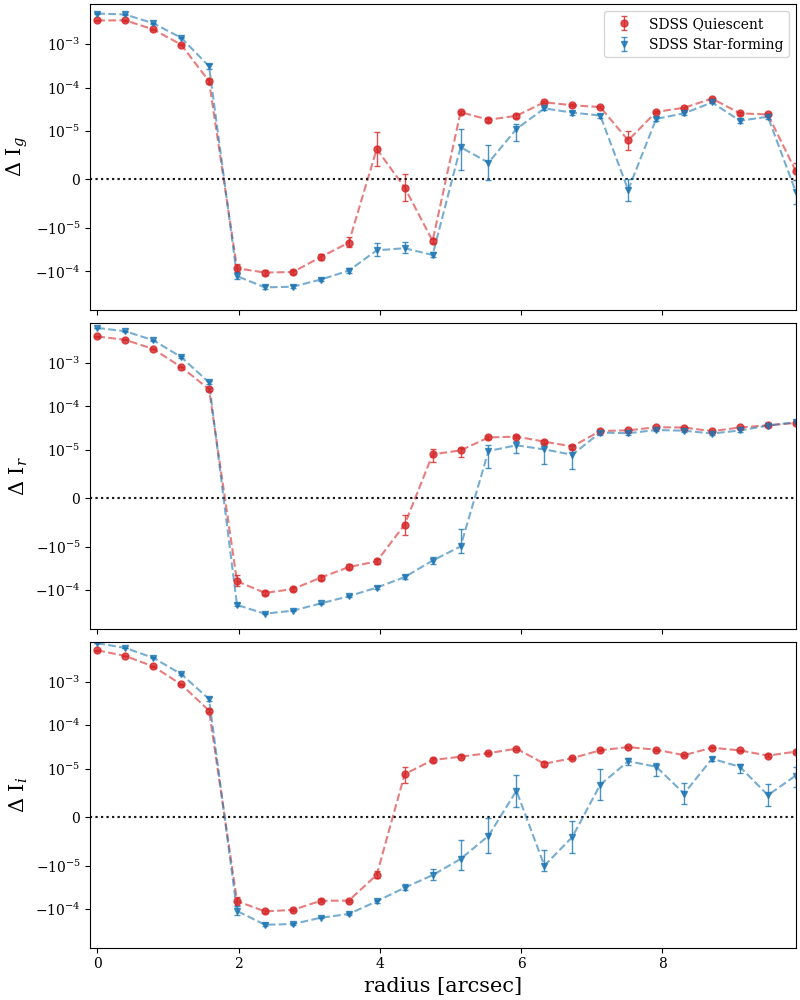}
    \caption{Difference between the surface brightness profiles of the SDSS low S/N E+A stacks and the comparison quiescent and star-forming stacks. The plots from top to bottom are $g$ band data (top), $r$ band data (middle), and $i$ band data (bottom). The red circles represent the quiescent stacks and the blue triangles represent the star-forming stacks, with the dashed line between points showing the average profile difference. The errors represent the standard error of the comparison profiles. A positive $\Delta I$ indicates the E+A profile is brighter at this radial bin compared to the comparison profile.}
    \label{fig:sdssdiff70}
\end{figure}

\begin{figure}
    \centering
    \includegraphics[width=\columnwidth]{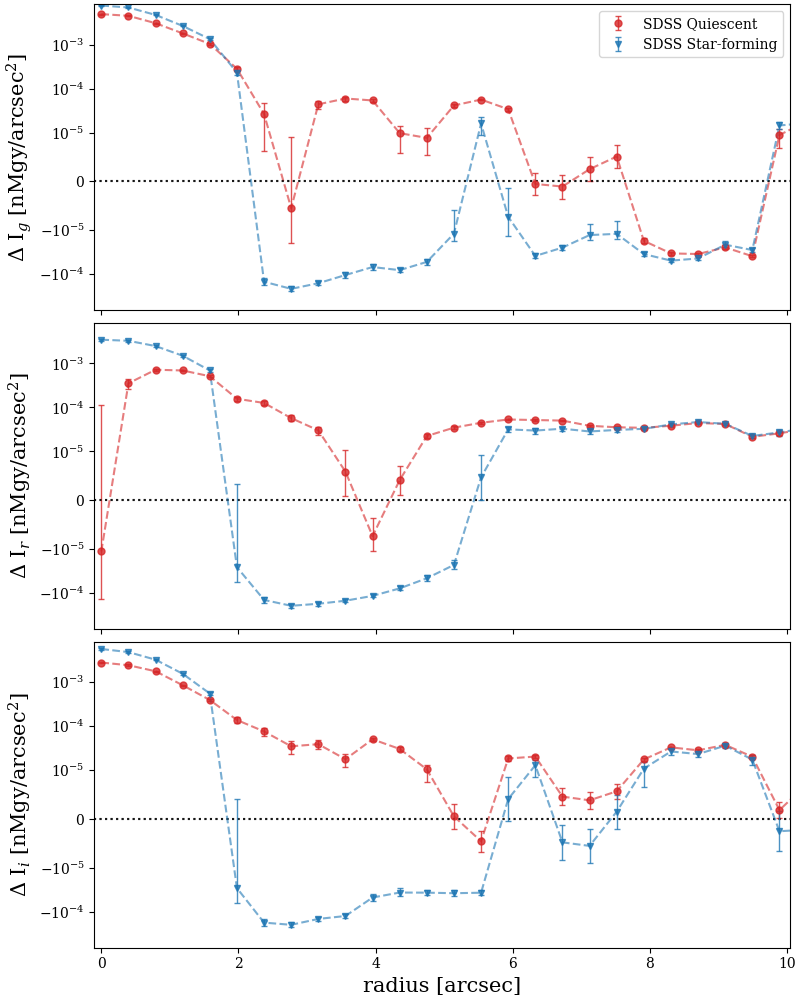}
    \caption{Difference between the surface brightness profiles of the SDSS high S/N E+A stacks and the comparison quiescent and star-forming stacks. The plots and key are the same as in Figure. \ref{fig:sdssdiff70}. A positive $\Delta I$ indicates the E+A profile is brighter at this radial bin compared to the comparison profile.}
    \label{fig:sdssdiff20}
\end{figure}

We quantise the differences between the E+A and comparison stacks in Figures. \ref{fig:sdssdiff70} and \ref{fig:sdssdiff20}. The plots highlight the brighter core in the E+A stacks compared to both the quiescent and star-forming as well as showing the dimmer intermediate radii ($2''<r<5''$). In the low S/N quiescent stacks, $\Delta I_{gri}>0$ when $r>5''$, meaning the E+A stacks have brighter extended structures compared to quiescents. In the high S/N stacks, we broadly see that same trends for a bright core and dimmer intermediate region for the E+A stacks, but at large radii the results differ significantly. We see that for the $g$ and $r$ bands, the comparison samples are brighter than the E+A galaxies at large radii. Additionally, the $i$ band also shows the comparison stacks peak slightly at the furthest radii ($r\approx10''$).

By calculating the outer-profile difference for thousands of bootstrapped instances, we find that the E+A galaxies in the low S/N stack exhibit a statistically higher amount enhanced outer light relative to both comparison samples. For the quiescent comparison, the outer-profile differences are $\Delta f_{\rm{outer}} = 0.346\pm0.106$, $0.371\pm0.100$, and $0.309\pm0.099$, yielding a significance of $3.27\sigma, 3.72\sigma$, and $3.13\sigma$ in the $g$, $r$, and $i$ bands respectively for the quiescent comparison difference. The star-forming comparisons show similar values with $\Delta f_{\rm{outer}} = 0.339\pm0.108$, $0.407\pm0.101$, and $0.276\pm0.103$, with significances of $3.13\sigma, 4.01\sigma$, and $2.70\sigma$ in the $gri$ bands. In all cases, 100\% of the bootstrap realisations produce a positive excess, providing strong evidence for enhanced light at these radii in the low S/N E+A stacks compared to both the quiescent and star-forming samples. 

Evidence for enhanced outer light in the high S/N E+A sample is only found in the $r$ and $i$ bands. Relative to the quiescent stacks, we measure $\Delta f_{\rm{outer}} = 0.386\pm0.094$, and $0.301\pm0.093$ for the $r$ and $i$ bands, with a $\approx3.64\sigma$ significance. Similar results are shown in the star-forming sample, $\Delta f_{\rm{outer}} = 0.445\pm0.098$, and $0.331\pm0.099$ for the $r$ and $i$ bands, with a $4.54\sigma$ and $3.34\sigma$ significance. All bootstrap realisations in the $r$ and $i$ bands shown a positive excess, suggesting that the high S/N E+A galaxies have more enhanced light at large radii than both comparison samples. In contrast, the $g$ band shows a negative difference, with $\Delta f_{\rm{outer}} = -0.018\pm0.092$, with $\approx74.5\%$ of realisations indicating a deficit rather than excess in the quiescent sample. In the star-forming sample, we measure $\Delta f_{\rm{outer}} = -0.042\pm0.096$ in the $g$ band, with $\approx63.2\%$ of realisations showing a negative difference. This suggests that there is more enhanced extended features in the comparison samples that are only found in the $g$ band. This could suggest that the tidal features in the E+A galaxies are not young and star forming, so won't appear in the $g$ band. Alternatively, dust obscuration could be hiding the blue light in the $g$ band but be present in the $r$ and $i$ bands.

Importantly, these differences must be interpreted in the context of the sample sizes, with the low S/N stack containing 57 galaxies, and the high S/N stack containing 17. The reduced sample size implies a greater susceptibility to stochastic effects and residual systematics at large radii. This apparent band dependant excess is more likely reflective of statistical variance, with reduced statistical sampling of faint outskirts and increased sensitivity to outliers.

The S\'ersic indices are systematically larger in all E+A stacks compared to both the quiescent and star-forming comparison samples (see Tables. \ref{tab:lowSNsersic results} \& \ref{tab:highSNsersic results}). This indicates that the E+A population is more centrally concentrated on average, with a steeper inner profile and more extended outer wings than the comparison galaxies. These $n$ values are consistent with structural change driven by a centrally focused evolutionary event such as a merger-driven event \citep{Yan2019}. This result differs from previous work in the literature where E+A systems occupy intermediate $n$ values between star-forming and quiescents \citep[e.g.][]{Sazonova2021}. This discrepancy may reflect differences in sample selection or be due to the stacking process, which can bias the S\'ersic fits toward higher central concentration in stacked analysis.

%%%%%%%%%%%%%%%%%%%%%%%%%%%%%%%%%%%%%%%%%%%%%%%%%%

\subsection{Legacy Survey Images}

\begin{table}
\caption{S\'ersic fitting parameters for the Legacy Survey low S/N E+A stack compared to the comparison Quiescent (Q) and Star-forming (SF). $r_e$ is the effective radius, $I_e$ is the surface brightness at $r=r_e$ and $n$ is the S\'ersic index. The error columns are found via bootstrapping 1000 times. The images for both the comparison and E+A stacks are extracted from DESI Legacy Survey DR10 and are of the same objects as in the SDSS image stacks.}
\label{tab:DES lowSNsersic results}
\begin{tabular}{llll}
 \hline
 & E+A Value & Q Value & SF Value \\ \hline
$g$ band & & & \\
$r_e {\rm{[arcsec]}}$        & 2.872$\pm$0.153 & 3.004$\pm$0.123 & 3.188$\pm$0.162 \\
$I_e {\rm{[nMgy/arcsec^2]}}$ & 0.103$\pm$0.011 & 0.094$\pm$0.008 & 0.132$\pm$0.014 \\
$n$                          & 1.733$\pm$0.256 & 1.819$\pm$0.187 & 1.11$\pm$0.074 \\ \hline
$r$ band & & & \\
$r_e {\rm{[arcsec]}}$        & 2.474$\pm$0.115 & 2.57$\pm$0.091 & 2.985$\pm$0.123 \\
$I_e {\rm{[nMgy/arcsec^2]}}$ & 0.463$\pm$0.038 & 0.42$\pm$0.032 & 0.335$\pm$0.021 \\
$n$                          & 1.629$\pm$0.178 & 1.756$\pm$0.117 & 1.112$\pm$0.065 \\ \hline
$i$ band & & & \\ 
$r_e {\rm{[arcsec]}}$        & 2.301$\pm$0.17 & 2.251$\pm$0.087 & 2.624$\pm$0.159 \\
$I_e {\rm{[nMgy/arcsec^2]}}$ & 0.815$\pm$0.064 & 0.694$\pm$0.053 & 0.492$\pm$0.043 \\
$n$                          & 1.789$\pm$0.213 & 1.875$\pm$0.213 & 1.241$\pm$0.149 \\ \hline
\end{tabular}
\end{table}

\begin{table}
\caption{S\'ersic fitting parameters for the Legacy Survey high S/N E+A stack compared to the comparison Quiescent (Q) and Star-forming (SF). $r_e$ is the effective radius, $I_e$ is the surface brightness at $r=r_e$ and $n$ is the S\'ersic index. The error columns are found via bootstrapping 1000 times. The images for both the comparison and E+A stacks are extracted from DESI Legacy Survey DR10 and are of the same objects as in the SDSS image stacks.}
\label{tab:DES highSNsersic results}
\begin{tabular}{llll}
 \hline
 & E+A Value & Q Value & SF Value \\ \hline
$g$ band & & & \\
$r_e {\rm{[arcsec]}}$        & 3.246$\pm$0.187 & 2.907$\pm$0.173 & 3.309$\pm$0.239 \\
$I_e {\rm{[nMgy/arcsec^2]}}$ & 0.271$\pm$0.031 & 0.195$\pm$0.033 & 0.220$\pm$0.027 \\
$n$                          & 1.139$\pm$0.122 & 1.332$\pm$0.106 & 0.941$\pm$0.136 \\ \hline
$r$ band & & & \\
$r_e {\rm{[arcsec]}}$        & 3.102$\pm$0.184 & 2.673$\pm$0.083 & 3.163$\pm$0.215 \\
$I_e {\rm{[nMgy/arcsec^2]}}$ & 0.763$\pm$0.062 & 0.636$\pm$0.062 & 0.477$\pm$0.063 \\
$n$                          & 1.237$\pm$0.087 & 1.399$\pm$0.094 & 0.965$\pm$0.11 \\ \hline
$i$ band & & & \\ 
$r_e {\rm{[arcsec]}}$        & 2.823$\pm$0.249 & 2.38$\pm$0.147 & 2.744$\pm$0.306 \\
$I_e {\rm{[nMgy/arcsec^2]}}$ & 1.327$\pm$0.065 & 0.929$\pm$0.119 & 0.626$\pm$0.106 \\
$n$                          & 1.284$\pm$0.163 & 1.643$\pm$0.191 & 1.165$\pm$0.203 \\ \hline
\end{tabular}
\end{table}

\begin{figure*}
    \centering
    \includegraphics[width=\textwidth]{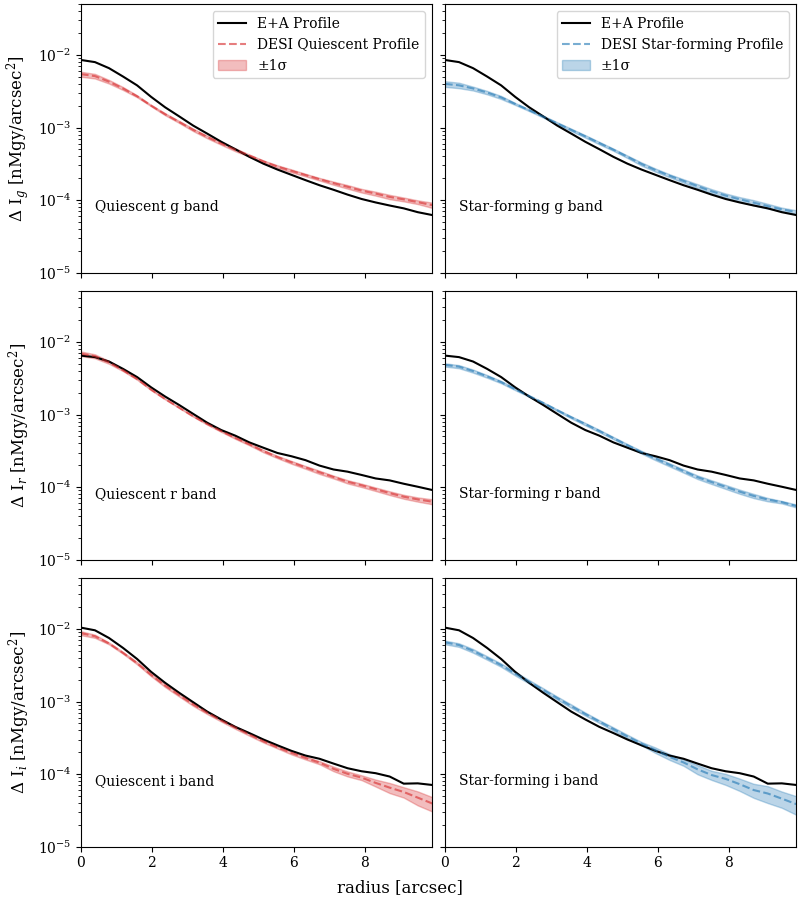}
    \caption{Surface brightness profiles of the Legacy Survey low S/N E+A stacks compared to the profiles of the quiescent (left) and star-forming (right) stacks. The key is the same as in Figure. \ref{fig:70comparestacks}. The E+A stacks typically have brighter central fluxes ($r<1''$) across both quiescent and star-forming stacks. In the $r$ and $i$ bands, the E+A stacks are brighter at larger radii than both comparison samples, however, in the $g$ band we can see that the comparison profiles are brighter at larger radii.}
    \label{fig:DES70comparestacks}
\end{figure*}

\begin{figure*}
    \centering
    \includegraphics[width=\textwidth]{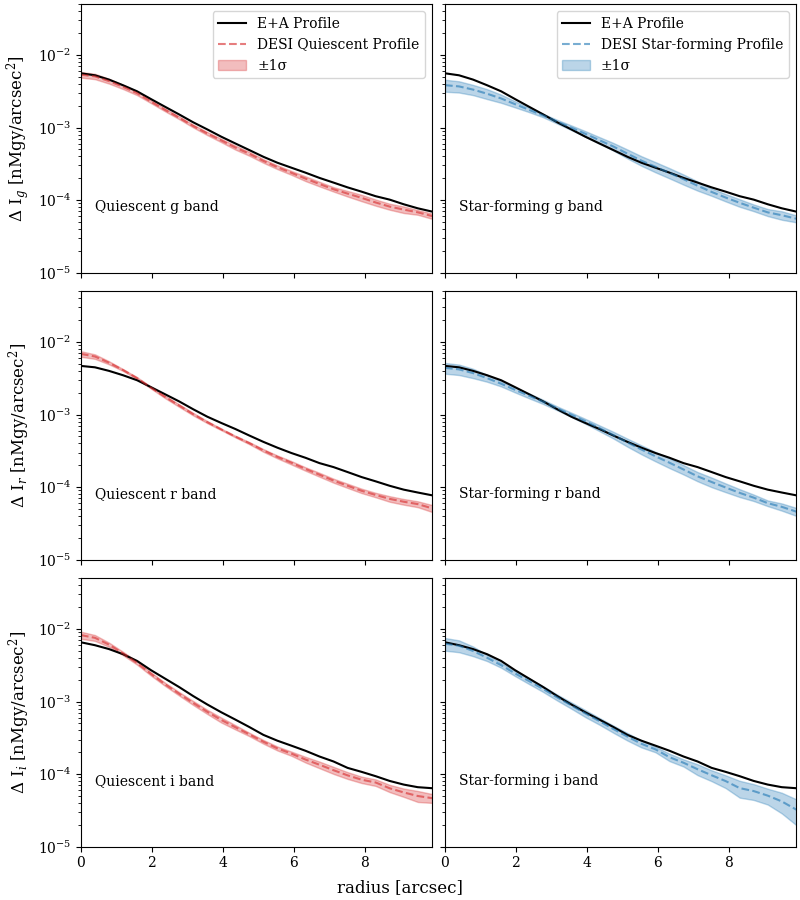}
    \caption{Surface brightness profiles of the Legacy Survey high S/N E+A stacks compared to the profiles of the quiescent (left) and star-forming (right) stacks. The key is the same as in Figure. \ref{fig:70comparestacks}. The E+A stacks in the $g$ band display slightly brighter central regions, but in the $r$ and $i$ bands the opposite is seen, with brighter central regions in the comparison stacks. All E+A stacks however are brighter than the comparison stacks are large radii.}
    \label{fig:DES20comparestacks}
\end{figure*}

The Legacy Survey radial profiles broadly follow the same trends as the SDSS radial profiles but do not have as severe difference in central brightness. In the $g$ band for the low S/N stacks, the brighter central emission in the E+A stacks is consistent with a centrally concentrated, relatively young stellar population that would contribute more strongly at bluer wavelengths. This is typical of E+A systems following a dissipative starburst or merger-driven inflow \citep{Zabludoffetal1996, Yang2004, French2021}. The  suppression of the $g$ band outskirts suggests a rapid decline in recent star formation in the outer regions. In contrast, the $r$ and $i$ band behaviour shows similar central intensities between E+A and comparisons stacks but brighter E+A outskirts. This indicates an outer structure dominated by older, redder stellar populations and/or diffuse low-surface-brightness components that are more effectively recovered in the Legacy Survey images. This would be physically consistent with a scenario in which that central starburst has faded, while the outer regions retains the tidal debris from a recent interaction. Similar extended faint structures are common in the literature, where merger remnants are seen more commonly in PSB galaxies with sufficiently deep imaging \citep{Pawlik2019, Sazonova2021, Verrico2023, DOnofrioetal2025}.

The Legacy Survey high S/N stack again follows a similar trend as the SDSS profiles, with all E+A stacks have brighter outskirts than the comparison stacks. Figure. \ref{fig:DES20comparestacks} shows the star-forming radial profiles are similar in shape to the E+A profiles, only diverging significantly at larger radii. The quiescent profiles also display similar shapes but they show brighter central regions compared to the E+A profile. This brighter region in the $r$ and $i$ bands indicates a strong population of old, redder stars, typically of a quiescent galaxy.

\begin{figure}
    \centering
    \includegraphics[width=\columnwidth]{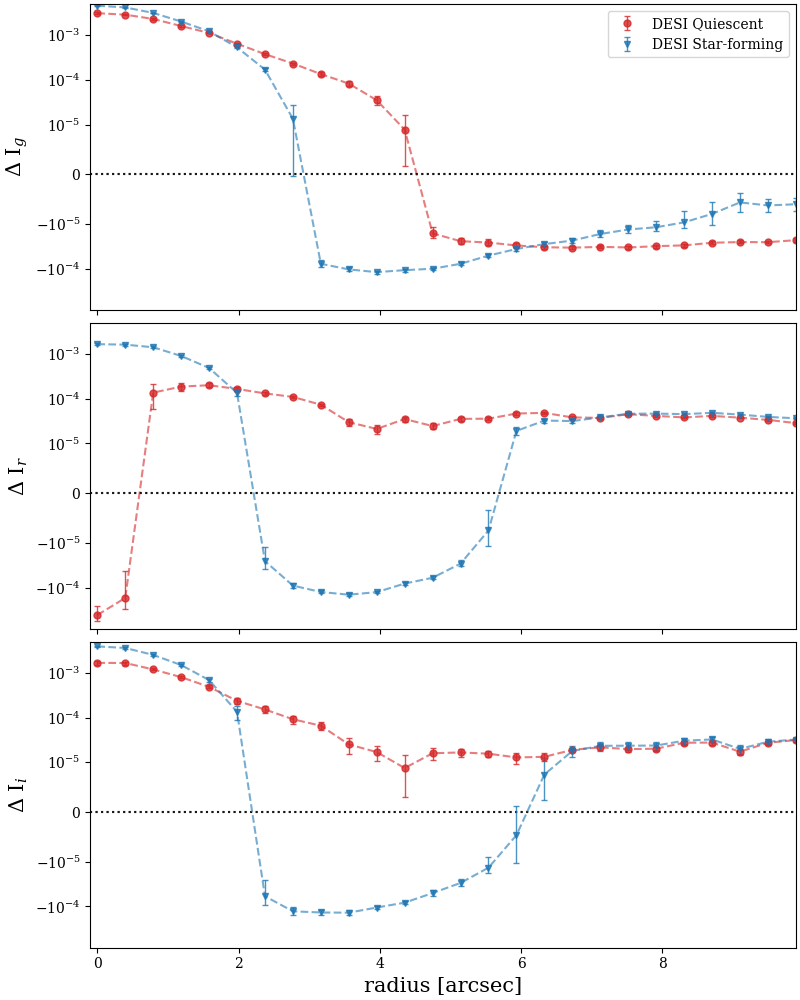}
    \caption{Difference between the surface brightness profiles of the Legacy Survey low S/N E+A stacks and the comparison quiescent and star-forming stacks. The plots and key are the same as in Figure. \ref{fig:sdssdiff70}. A positive $\Delta I$ indicates the E+A profile is brighter at this radial bin compared to the comparison profile.}
    \label{fig:desdiff70}
\end{figure}

\begin{figure}
    \centering
    \includegraphics[width=\columnwidth]{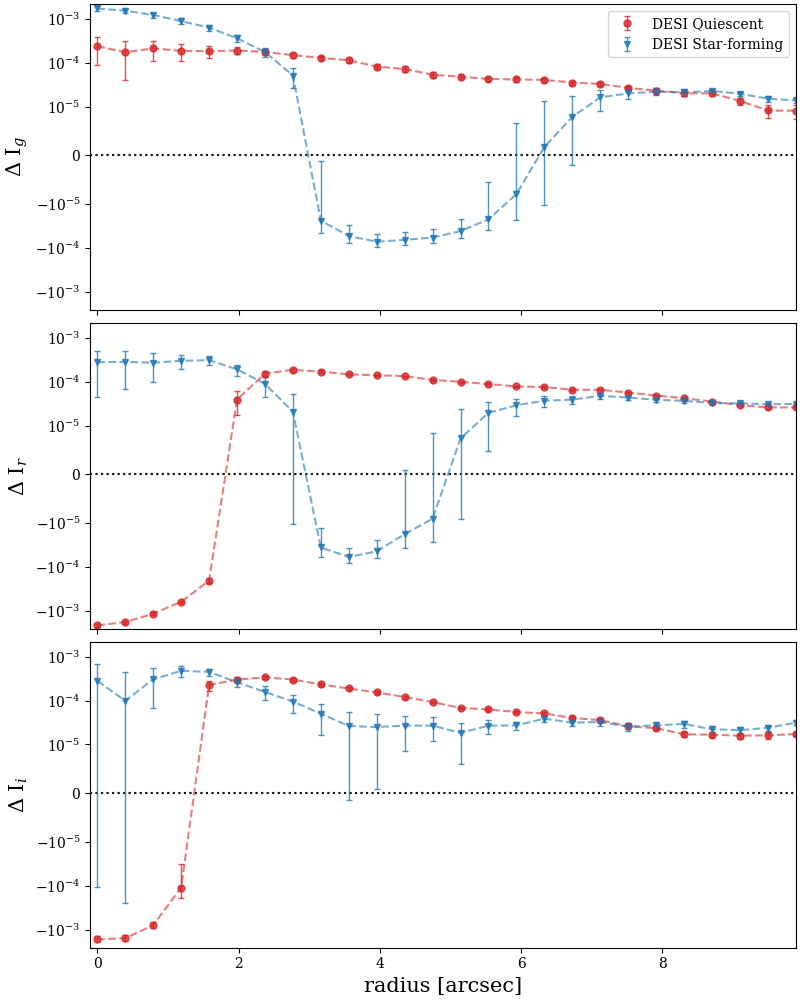}
    \caption{Difference between the surface brightness profiles of the Legacy Survey high S/N E+A stacks and the comparison quiescent and star-forming stacks. The plots and key are the same as in Figure. \ref{fig:sdssdiff70}. A positive $\Delta I$ indicates the E+A profile is brighter at this radial bin compared to the comparison profile.}
    \label{fig:desdiff20}
\end{figure}

The difference between surface brightness profiles is again quantised and plotted in Figures. \ref{fig:desdiff70} and \ref{fig:desdiff20}. The Legacy Survey E+A stacks don't typically show the bright core seen in the SDSS E+A stacks, except compared to the quiescent $r$ band in the low S/N stack and the quiescent $r$ and $i$ bands for the high S/N stack. The intermediate bulge is also seen more predominantly in the star-forming stacks and not the quiescent stacks. This shows a more exponential disk, common in star-forming spirals. The E+A profiles are brighter a large radii ($r>6''$) in all bands and stacks except the star-forming and quiescent low S/N $g$ band. This suggests again that the $g$ band outskirts have been suppressed in the E+A stack, indicating a rapid decline of star formation \citep{Yang2004, French2021}.

Applying the same outer profile difference to the Legacy Survey stacks, we measure a positive difference between the low S/N E+As and the quiescent comparisons in the $r$ and $i$ bands, with $\Delta f_{\rm{outer}} = 0.308\pm0.099$, and $0.268\pm0.092$, with a significance of $3.11\sigma$ and $2.91\sigma$.  For the star-forming comparisons, we measure a significance $2.87\sigma, 3.59\sigma$, and $3.11\sigma$ across the $g,r$ and $i$ bands. In contrast, the $g$ band shows a negative difference in low S/N - quiescent comparison, with $\Delta f_{\rm{outer}} = 0.0501\pm0.109$. This suggests quiescent galaxies have brighter extended structure than both E+As and star-formers in the $g$ band for this sample. Dust obscuration could account for this finding, or the tidal features in the quiescent galaxies could be close to the detection threshold, excluding them in the $r$ and $i$ bands.

The difference between the high S/N E+As and the comparisons samples are often positive. We measure $\Delta f_{\rm{outer}} = 0.258\pm0.083$, $0.224\pm0.076$, and $0.165\pm0.063$ in the quiescent $gri$ bands, and $\Delta f_{\rm{outer}} = 0.251\pm0.082$, $0.279\pm0.089$, and $0.233\pm0.091$ in the star-forming $gri$ bands. These differences have a significance of $3.11\sigma, 2.95\sigma$, and $2.62\sigma$ for the quiescent comparisons, and  $3.06\sigma, 3.13\sigma$, and $2.56\sigma$ for the star-forming comparisons.

The Legacy Surveys improved depth compared to SDSS is also reflective in the retrieved S\'ersic parameters, with the apparent extended wings lowering the $n$ values in line with literature \citep[e.g.][]{Sazonova2021}. We see in Tables. \ref{tab:DES lowSNsersic results} and \ref{tab:DES highSNsersic results} that $n$ is more intermediate than the prior SDSS values. The E+A indices lie above the star-forming indices and below the quiescent indices, suggesting a more intermediate morphology between elliptical and spiral. These results likely come from the low-surface-brightness features being recovered more effectively in the Legacy Survey images. It is also reflective of the sensitivity of single-component S\'ersic fits to data depth and the deeper image reduces biases that can artificially inflate S\'ersic indices \citep{Haussler2007}.

\begin{figure}
    \centering
    \includegraphics[width=\columnwidth]{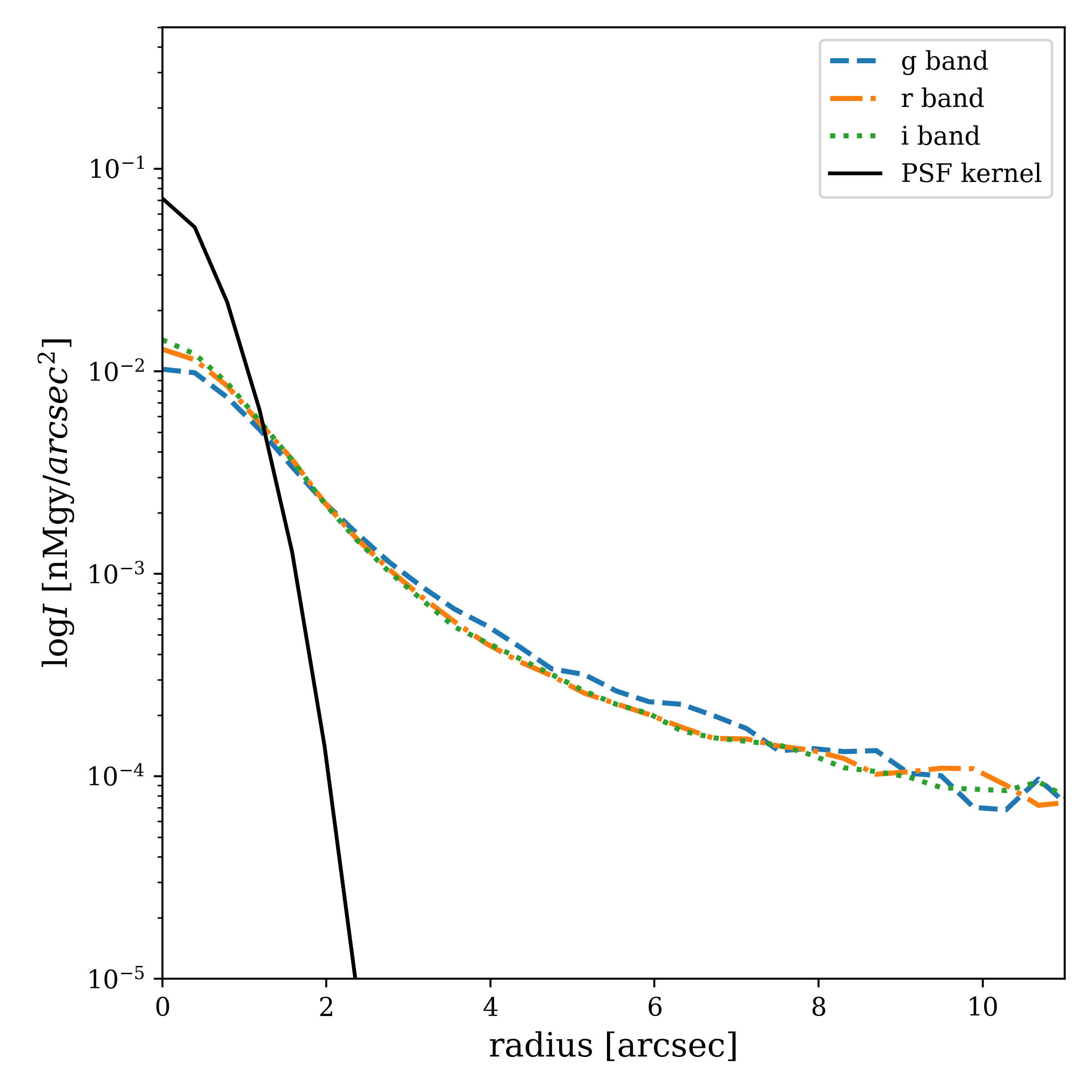}
    \caption{SDSS low S/N E+A radial profiles compared to the PSF kernel used to convolve them. The PSF is shown by the solid black line and the $gri$ bands are shown by the coloured dashed, dash-dot, and dotted lines respectively. The PSF decreases steeply with radius and as such the PSF wings should not be affecting the E+A surface brightness profiles.}
    \label{fig:psfcheck}
\end{figure}

\section{Discussion}

Small-scale interactions are considered a common driving mechanism for starting and halting star-formation in post-starburst galaxies \citep{Pawliketal2018, Matharuetal2020, Lietal2023}. Faint tidal features not traceable by shape asymmetry could also be signs of major mergers, with these feature fading on 200Myr timescales \citep{Pawlik2019}. Work like \citet{Verrico2023} show through visual classification that a disturbed morphology is more common in PSB galaxies compared to both star-forming and quiescent galaxies. We find however, that many of our single E+A galaxies do not show a clear extended structure due to poorly resolved images. \citet{Verrico2023} notes this limitation and utilises a subsample of PSB galaxies that overlap with the Hyper Suprime-Cam (HSC) Survey PDR 3 \citep{Aiharaetal2018, Aiharaetal2022}. These images are much deeper and have higher resolution that SDSS and make visual classifications of galaxies more reliable. Modern surveys and tools are also in the process of improving upon these results, such as the Legacy Survey of Space and Time (LSST), and we anticipate that the extended structure of E+A will be even better resolved. 

Several pieces of literature have analysed the morphologies and S\'ersic parameters of E+As compared to comparative quiescent and star-forming galaxies. Notably, \citet{Sazonova2021} analyses the morphology of shocked post-starburst galaxies (SPOGs) from the Hubble Space Telescope (HST). They find their SPOGs have an intermediate morphology between disk- and bulge-dominated galaxies, with predominantly red bulges. They also find that a majority of the SPOGs have disturbed morphologies, in line with other works \citep{Pawliketal2018, Verrico2023, DOnofrioetal2025}. Many of these works utilise deeper imaging to uncover these faint sub-structures, with \citet{Pawliketal2018} noting that more resolved spectroscopy and upcoming datasets will be crucial for future analysis. 

We aim to uncover the faint sub-structures not seen through means such as shape asymmetry \citep{Pawliketal2016} or other non-parametric quantities via mean image stacking. Through this image stacking we observe surface brightness profiles that suggests faint extended structure in many of the SDSS E+A stacks and Legacy Survey E+A stacks when compared to quiescent and star-forming stacks. The SDSS image stacks show promising results, with extended structure being apparent in the larger low S/N stack. This structure is less apparent in the smaller high S/N stack, which could be due to sample size increasing susceptibility to outliers and residual errors. The Legacy Survey stacks reinforce the presence of extended structure in both the E+A samples, likely due to the greater depth revealing faint structure that wasn't present in the SDSS imaging. Faint surface brightness structures at large radii are a direct signature of possible mergers or small-scale galaxy-galaxy interactions \citep{Lotzetal2008, Pawliketal2016, Verrico2023}. 

The E+As exhibit a significant excess of low-surface-brightness light at large radii compared to the comparison non-E+As, consistent with enhanced tidal debris. This tidal debris could be an indicator of small-scale interactions such as fly-bys or harassment or, it could be the faded structure of a major merger event in an older E+A galaxy \citep{Pawliketal2018}. We suggest the extended structure in our E+A galaxies is a result of these interactions, and as such, small-scale interactions and/or mergers are a common mechanism behind quenching in E+A galaxies. These results are well aligned with the literature, where galaxy evolution is likely driven by minor mergers \citep{Newman2012, Pawliketal2018, Sazonova2021, Zhang2024, Haryana2025}.

Other than sky background residuals affecting the results, we also analysed the effect the PSF has on the results and whether the extended structure is simply extended PSF wings. We therefore extract stars from the E+A plates used and calculate the PSF of the images. Plotting this PSF, shown in Fig. \ref{fig:psfcheck}, we can see that there is a significant decrease in PSF as radius increases. This indicates that large PSF wings are not the cause of the bright extended structure.

The S\'ersic results for the SDSS stacks do not support the intermediate morphology that is typical of E+A galaxies \citep{Sazonova2021}. The Legacy Survey stacks are in line with the literature, however, having intermediate morphologies with $n_g=1.73\pm0.26$, $n_r=1.63\pm0.18$, and $n_i=1.79\pm0.21$ in the low S/N E+A stacks and $n_g=1.14\pm0.12$, $n_r=1.24\pm0.09$, and $n_i=1.28\pm0.16$ in the high S/N stacks. A plausible explanation for the discrepancy between the two stacks could be the sensitivity to low-surface-brightness structure and associated systematic effects. SDSS imaging is shallower, so the faint extended components might be more susceptible to suppression due to systematic problems and leading to increased S\'ersic fits. In addition, limitations in SDSS sky subtraction may lead to an over-subtraction of extended low-surface-brightness emission \citep{Blanton2011}, artificially truncating outer profiles. In deeper surveys, the same structures in the E+A galaxies will be detected at a higher S/N, making them less vulnerable to being washed out by the stacking process \citep{Li2022}, resulting in less centrally dominated S\'ersic indices. Further, even deeper surveys such as the Rubin Observatory Legacy Survey of Space and Time (LSST) could provide us with more resolved faint structure, allowing us to probe the disturbed morphologies of E+A and PSB galaxies more effectively.

\begin{figure}
    \centering
    \includegraphics[width=\columnwidth]{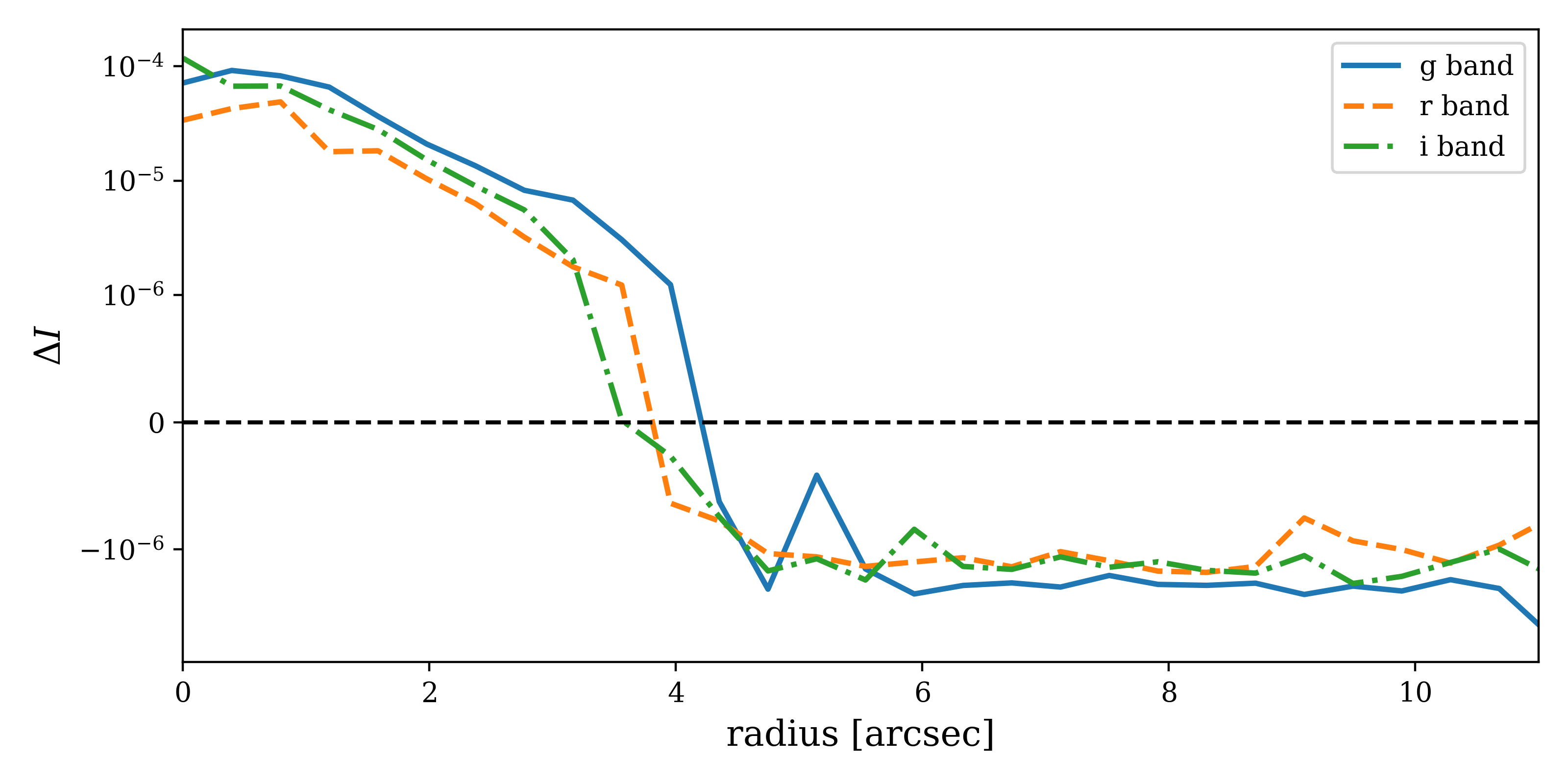}
    \caption{Difference plot of the SDSS low S/N E+A stack - E+A companion stack. The solid blue line is the $g$ band difference, orange dashed line the $r$ band difference, and the green dot-dash line the $i$ band difference. A positive difference indicates that the standard E+A stack has a brighter central region than that of the companion E+A stack. We find that the companion stack has a brighter extended region but both differences are an order of magnitude lower than the differences between the E+A and quiescent and star-forming comparisons.}
    \label{fig:companiondiff}
\end{figure}

The manual removal of contaminated images may introduce a selection bias if physically associated companion galaxies are preferentially excluded from the E+A sample. To test this possibility, we constructed additional stacks that retained plausible companions and repeated the analysis. These companion-inclusive stacks display marginally brighter outskirts and slightly reduced central surface brightness relative to the standard E+A stacks, indicating that nearby companions can contribute to the measured extended light profile. We display the difference between the low S/N SDSS image stacks of the standard and companion E+A galaxies in Figure. \ref{fig:companiondiff}. We note that the high S/N sample doesn't change as the PSF threshold will exclude the companion galaxy. The resulting change is approximately one order of magnitude smaller than the excess outer light observed between the E+A and comparison stacks. This suggests that although the image-cleaning procedure may influence the precise amplitude of the extended emission, the principal result of enhanced outer structure in the E+A population remains robust.

While the enhanced outer light observed in the E+A stacks is consistent with tidal debris arising from mergers or other small-scale interactions, alternative explanations remain possible. Extended stellar discs or haloes, unresolved companion galaxies, residual centring uncertainties, PSF scattering, and incomplete matching of the comparison samples could also contribute to the observed excess at large radii. We have taken care to minimise these effects through masking, PSF matching, redshift scaling, comparison-sample selection, and robustness tests, but they cannot be entirely excluded. As such, our interpretation should be viewed as evidence that is consistent with interaction-driven origins rather than as a unique explanation for the observed extended structure.

\section{Conclusion}

We have investigated the extended structure of 57 E+A galaxies selected from GAMA DR4 using image stacking techniques applied to both SDSS and DESI Legacy Survey imaging. By constructing stacked surface-brightness profiles and fitting S\'ersic models, we find that E+A galaxies exhibit systematically enhanced low-surface-brightness emission at large radii when compared to carefully matched quiescent and star-forming galaxy samples. This excess outer light is most prominent in the larger low S/N sample and is further reinforced by the deeper Legacy Survey data, which recover faint structures more effectively than SDSS imaging.

The observed enhancement of outer light is consistent with the presence of tidal debris and disturbed stellar structures associated with galaxy interactions. Combined with previous studies that identify merger signatures and disturbed morphologies in PSB systems, our results support the scenario in which interactions, such as mergers, flybys, or harassment, play a significant role in triggering and/or quenching star formation in E+A galaxies \citep{Pawlik2019, Verrico2023, DOnofrioetal2025}. Furthermore, the intermediate S\'ersic indices derived from the deeper Legacy Survey images are consistent with E+A galaxies occupying a transitional morphological state between star-forming discs and quiescent spheroids \citep{Sazonova2021}.

Future deep, wide-area surveys such as Rubin LSST will provide substantially improved sensitivity to faint tidal features and diffuse stellar structures. These observations will enable larger statistical studies of PSB galaxies and help clarify the relative importance of mergers and small-scale interactions in driving rapid galaxy evolution.

%%%%%%%%%%%%%%%%%%%%%%%%%%%%%%%%%%%%%%%%%%%%%%%%%%

\section*{Acknowledgements}

We thank the anonymous referee for valuable feedback that has improved the robustness and reliability of these results.
This research made extensive use of {\sc{ipython}} \citep{PER-GRA:2007} and the following {\sc{python}} packages: astropy\footnote{\url{www.astropy.org/}}, a community-developed core Python package and an ecosystem of tools and resources for astronomy \citep{astropy:2013, astropy:2018, astropy:2022}; astroquery\footnote{\url{astroquery.readthedocs.io/en/latest/}}, a coordinated package for astropy \citep{Astroquery}; astropop\footnote{\url{https://astropop.readthedocs.io/en/latest/}}, the ASTROnomical POlarimetry and Photometry pipeline, developed for work with IAGPOL and SPARC4 polarimeters at Observatório Pico dos Dias (Brazil) \citep{astropop}; photutils\footnote{\url{https://photutils.readthedocs.io/en/stable/}}, a python library that provides commonly-used tools and key functionality for detecting and performing photometry of astronomical sources \citep{photutils};
scikit-learn\footnote{\url{www.scikit-learn.org/}} \citep{Sklearn}; and pysersic\footnote{\url{https://pysersic.readthedocs.io/en/latest/}}, a package for fitting S\'ersic profiles to astronomical images using Bayesian inference \citep{pysersic}.

%%%%%%%%%%%%%%%%%%%%%%%%%%%%%%%%%%%%%%%%%%%%%%%%%%
\section*{Data Availability}

This work made extensive use of the GAMA DR4 data. GAMA is a joint European-Australasian project based around a spectroscopic campaign using the Anglo-Australian Telescope. The GAMA input catalogue is based on data taken from the Sloan Digital Sky Survey and the UKIRT Infrared Deep Sky Survey. Complementary imaging of the GAMA regions is being obtained by several independent survey programmes including GALEX MIS, VST KiDS, VISTA VIKING, WISE, Herschel-ATLAS, GMRT and ASKAP providing UV to radio coverage. GAMA is funded by the STFC (UK), the ARC (Australia), the AAO, and the participating institutions. The GAMA website is \hyperlink{https://www.gama-survey.org/}{https://www.gama-survey.org/}. We make specific use of SpecLinesSFR \citep{Gordonetal2017}, which compiles a significant amount of spectroscopic data and EW measures. We further use GKVScienceCat \citep{Bellstedtetal2020}, which compiles the main survey selection including redshifts and photometry data. We utilise SpecCat and ApMatchedPhotom \citep{Liskeetal2015} for extracting the spectra plots and for further photometry measures. We make use of StellarMasses \citep{Tayloretal2011} for creating our comparison galaxy image stacks.

This research used data obtained with the Dark Energy Spectroscopic Instrument (DESI). DESI construction and operations is managed by the Lawrence Berkeley National Laboratory. This material is based upon work supported by the U.S. Department of Energy, Office of Science, Office of High-Energy Physics, under Contract No. DE–AC02–05CH11231, and by the National Energy Research Scientific Computing Center, a DOE Office of Science User Facility under the same contract. Additional support for DESI was provided by the U.S. National Science Foundation (NSF), Division of Astronomical Sciences under Contract No. AST-0950945 to the NSF’s National Optical-Infrared Astronomy Research Laboratory; the Science and Technology Facilities Council of the United Kingdom; the Gordon and Betty Moore Foundation; the Heising-Simons Foundation; the French Alternative Energies and Atomic Energy Commission (CEA); the National Council of Humanities, Science and Technology of Mexico (CONAHCYT); the Ministry of Science and Innovation of Spain (MICINN), and by the DESI Member Institutions: \hyperlink{www.desi.lbl.gov/collaborating-institutions}{www.desi.lbl.gov/collaborating-institutions}. The DESI collaboration is honoured to be permitted to conduct scientific research on I’oligam Du’ag (Kitt Peak), a mountain with particular significance to the Tohono O’odham Nation. Any opinions, findings, and conclusions or recommendations expressed in this material are those of the author(s) and do not necessarily reflect the views of the U.S. National Science Foundation, the U.S. Department of Energy, or any of the listed funding agencies.


\bibliographystyle{mnras}
\bibliography{00reference}

%%%%%%%%%%%%%%%%%%%%%%%%%%%%%%%%%%%%%%%%%%%%%%%%%%

%%%%%%%%%%%%%%%%% APPENDICES %%%%%%%%%%%%%%%%%%%%%

\appendix

%%%%%%%%%%%%%%%%%%%%%%%%%%%%%%%%%%%%%%%%%%%%%%%%%%

% Don't change these lines
\bsp	% typesetting comment
\label{lastpage}
\end{document}